%% file: main.tex
\documentclass{article}

\pdfoutput=1

\usepackage{arxiv}

\usepackage[utf8]{inputenc} 
\usepackage[T1]{fontenc}    
\usepackage{hyperref}       
\usepackage{url}            
\usepackage{booktabs}       
\usepackage{amsfonts}       
\usepackage{nicefrac}       
\usepackage{microtype}      
\usepackage{lipsum}		
\usepackage{graphicx}
\usepackage{natbib}
\usepackage{doi}
\usepackage{amsmath}
\usepackage{multirow}

\usepackage[ruled,linesnumbered]{algorithm2e} 

\SetAlFnt{\small}
\SetAlCapFnt{\small}
\SetAlCapNameFnt{\small}
\SetAlCapHSkip{0pt}

\usepackage[braket, qm]{qcircuit}

\usepackage{subfigure}

\usepackage{footnote}

\newcommand{\settAlg}[1]{\texttt{#1}}
\newcommand{\QTrace}{\texttt{QTrace()}\,}
\newcommand{\QSearch}{\texttt{QSearch()}\,}
\newcommand{\QOccluded}{\texttt{QOccluded()}\,}
\newcommand{\TracePass}{\texttt{TracePass()}\,}
\newcommand{\DirectPass}{\texttt{DirectPass()}\,}
\newcommand{\RenderScene}{\texttt{RenderScene()}\,}
\newcommand{\NeighOpt}{\texttt{NeighOpt()}\,}

\title{Towards Quantum Ray Tracing}

\author{
{Luís Paulo Santos}\\
{Universidade do Minho \&}\\
{INESC TEC}\\
{Braga}\\
{Portugal}
\And 
{Thomas Bashford-Rogers}\\
{Department of Computer Science and Creative Technologies}\\
{University of the West of England}\\
{Bristol}\\
{UK}
\And
{João Barbosa}\\
{Universidade do Minho \&}\\
{INESC TEC}\\
{Portugal}
\And
{Paul Navrátil}\\
{Texas Advanced Computing Center}\\
{University of Texas at Austin}\\
{USA}
}

\begin{document}
\maketitle

\begin{abstract}
Rendering on conventional computers is capable of generating realistic imagery, but the computational complexity of these light transport algorithms is a limiting factor of image synthesis. Quantum computers have the potential to significantly improve rendering performance through reducing the underlying complexity of the algorithms behind light transport. This paper investigates hybrid quantum-classical algorithms for ray tracing, a core component of most rendering techniques. Through a practical implementation of quantum ray tracing in a 3D environment, we show quantum approaches provide a quadratic improvement in query complexity compared to the equivalent classical approach. Based on domain specific knowledge, we then propose algorithms to significantly reduce the computation required for quantum ray tracing through exploiting image space coherence and a principled termination criteria for quantum searching. We show results for both Whitted style ray tracing, and for accelerating ray tracing operations when performing classical Monte Carlo integration for area lights and indirect illumination.
\end{abstract}

\keywords{quantum computing, ray tracing, quadratic query complexity quantum advantage}

\maketitle

\input{introduction}

\input{QuantumSearching}

\input{RelatedWork}


\input{QRT-Operator}

\input{QRT-Results}

\input{QRT-Optimizations}

\input{BeyondWRT}

\input{Discussion}

\input{Conclusion}

\section*{Acknowledgments}

This research was funded in part by the Intel Graphics and Visualization Institute of eXcellence at the Texas Advanced Computing Center.

\bibliographystyle{ACM-Reference-Format}
\bibliography{QRenderer}

\appendix
\input{QuantumComputing}

\input{QRT-algorithms-appendix}

\end{document}

%% file: introduction.tex
\section{Introduction}


Quantum computing, originally proposed by Benioff \cite{benioff1980computer} and Feynmann \cite{Feynman1982}, and developed into the quantum Turning Machine  \cite{deutsch1985quantum}, has many well known applications, from simulation of quantum systems \cite{lanyon2010towards} to factoring primes \cite{Shor1997}. These algorithms typically exploit the large space spanned by a quantum system, and that computations can performed simultaneously on this space, leading to an advantage over classical systems. However, both the quantum nature and the size of this space leads to practical problems when applying these concepts to rendering and graphics problems.

Existing quantum hardware \cite{Preskill2018} has recently become practical for small problems \cite{IBMQ}. However, most hardware implementations of quantum computers at the time of writing have limited \textit{quantum volume}, a metric \cite{cross2019validating} which describes the capability of quantum computers to solve problems. This leads to the machines reliably running very small programs before errors dominate the results. Quantum algorithms can also be simulated on classical computers, however this has exponentially large time and/or space requirements: each added quantum bit (qubit) doubles the time and/or the memory required for the simulation.

Therefore, in this paper we propose an approach which builds towards the goal of a fully quantum rendering system. However, directly simulating a modern, complicated 3D environment in a quantum computer is likely to require a significant increase in the quantum volume before this is practical. On the contrary, it is possible to simulate quantum hardware on classical computers and both develop quantum algorithms and gain an understanding of their performance without the constrains of a real machine.

In this paper we focus on ray tracing as it is the underlying algorithm behind most rendering algorithms. We primarily focus on the development and analysis of quantum searching algorithms for tracing primary, shadow and indirect rays in a 3D environment. Based on certain assumptions, we show that there is a computational advantage when compared to a classical approach without spatial data structures. While naive classical ray tracing has $\mathcal{O}(N)$ complexity in terms of the number of intersection evaluations, quantum algorithms have $\mathcal{O}(\sqrt{N})$ complexity, with $N$ being the number of geometric primitives in the scene. A quadratic advantage in query (intersections evaluation) complexity is thus achieved, which represents a major gain for large $N$.

While the proposed quantum algorithm leads to improvement over a classical approach, we observe that there is scope for many optimisations for the specific case of rendering. We propose two such optimisations leading to a hybrid quantum-classical algorithm. The first is to exploit image space coherence to reduce noise resulting from the non-deterministic nature of quantum systems, and the second is a novel principled termination criteria for the quantum searching algorithm which substantially decreases the amount of computation required to generate images.

\begin{figure}[tp]
\begin{center}
	\subfigure[Reference classical rendering: $\approx 2678 K$ intersections evaluated for 41842 rays (64 per ray).]{
	    \label{fig:C-Qornell064-WH128}
        \includegraphics[width=0.25\textwidth]{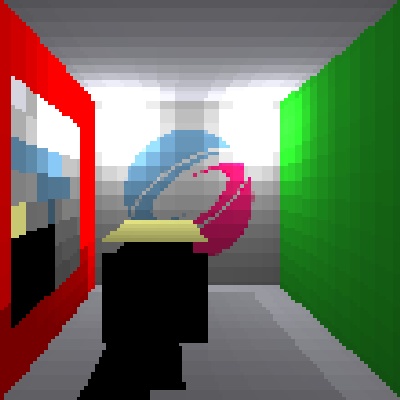}
    } \hspace{1cm}
	\subfigure[Non-optimized quantum rendering: $\approx 1366 K$ intersection evaluations (33.6 per ray).]{
	    \label{fig:Q_Qornell064_128x128_MP4}
        \includegraphics[width=0.25\textwidth]{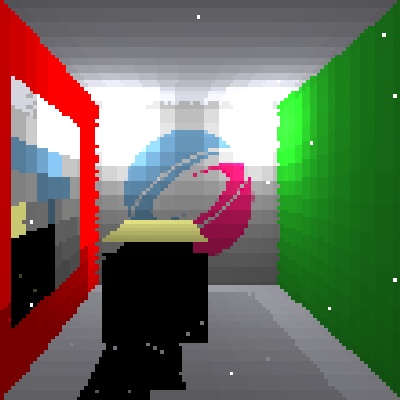}
    } \hspace{1cm}
    \subfigure[Optimized quantum rendering: $\approx 896 K$ intersection evaluations (22.1 per ray).]{
	    \label{fig:Q_Qornell064_128x128_NN_SN-teaser}
        \includegraphics[width=0.25\textwidth]{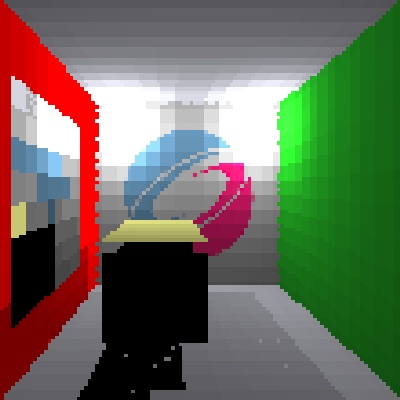}
    } 
    \caption{64 geometric primitives Qornell Box rendered at 128x128 resolution. This illustrates the main outcome of this work: the number of intersections required by our proposed quantum algorithm is shown to be less than classically required. However, the probabilistic nature of measuring intersections leads to visible artifacts (as can be seen in the artifacts in the middle image), so we propose a correction method which both resolves these errors, and decreases the number of intersection evaluations.}
\label{fig:QornellOverview}
\end{center}
\end{figure}

To summarize, the contributions of this paper are:

\begin{itemize}
    \item A hybrid quantum-classical ray tracer using quantum searching algorithms,
    \item A thorough analysis of the performance of quantum algorithms for ray tracing compared to classical ray tracing,
    \item A hybrid quantum-classical algorithm for optimizing performance though exploiting image space coherence and principled stopping criteria for quantum searching,
    \item Further examples of the use of quantum ray tracing such as evaluating visibility for Monte Carlo integration of area light sources and indirect illumination.
\end{itemize}

This is, to the extent of the authors' knowledge, the first hybrid quantum-classical ray tracing system for 3D environments to be demonstrated, confirming a quadratic advantage in a simulated environment, as predicted by the theory.



%% file: QuantumSearching.tex
\section{Quantum Searching}
\label{sec:QuantumSearching}

We start by introducing quantum searching over an unstructured database, which is a core building block of our approach. This section relies on some knowledge of quantum computing, and we present a short introduction to the fundamentals of quantum computing in appendix \ref{sec:IntroductionToQuantumComputing}.


Consider a function $f: \{0,1\}^n \rightarrow \{0,1\}$, or equivalently, in decimal notation, $f: \{0, 1,  \ldots N-1\} \rightarrow \{0,1\}$, where $N=2^n$ and $X=\{0, 1,  \ldots N-1\}$ denotes the set of all integers $\{i | 0\leq i \leq N-1\}$. Define f(x) as
\begin{equation}
  f(x) =
  \left \{
  \begin{array}{ll}
    1 & \text{if $x$ is a solution} \\
    0 & \text{otherwise.} 
  \end{array}
  \right.
  \label{eq:solutionIndicator}
\end{equation}
Searching can be described as the problem of finding a value $x \in X$ such that $f(x)=1$. 

If nothing is known about the structure of $f()$ and if there are $t$ solutions (i.e., $t$ different $x$ values $\in X$ such that $f(x)=1$), then a classical approach will find a solution after $N/(t+1)$ evaluations of $f()$ on average and $N-t+1$ such evaluations in the worst case. 

Let $\widehat{\mathcal{A}}$, referred to as the oracle, be the quantum operator implementing $f()$, such that 
\begin{equation}
\widehat{\mathcal{A}}|0\rangle^{\otimes n}|0\rangle = \frac{1}{\sqrt{N}} \sum_{i=0}^{N-1} |i\rangle |f(i)\rangle
\label{eq:GroverOracle}
\end{equation}
$\widehat{\mathcal{A}}$ sets the output qubit to $|1\rangle$ when the value $|i\rangle$ represented by the $n$ state qubits is a solution for $f()$. A single evaluation of the oracle computes the value of $f()$ for all points in $X$ -- quantum parallelism; however, these cannot be measured. Quantum searching enables finding, with high probability, a solution for the search problem with $\mathcal{O}(\sqrt{N/t})$ evaluations of $\widehat{\mathcal{A}}$ and its inverse, therefore providing a quadratic advantage over the classical result. This is achieved with Grover's algorithm \cite{Grover1996}.


\begin{figure*}
\includegraphics[scale=0.9]{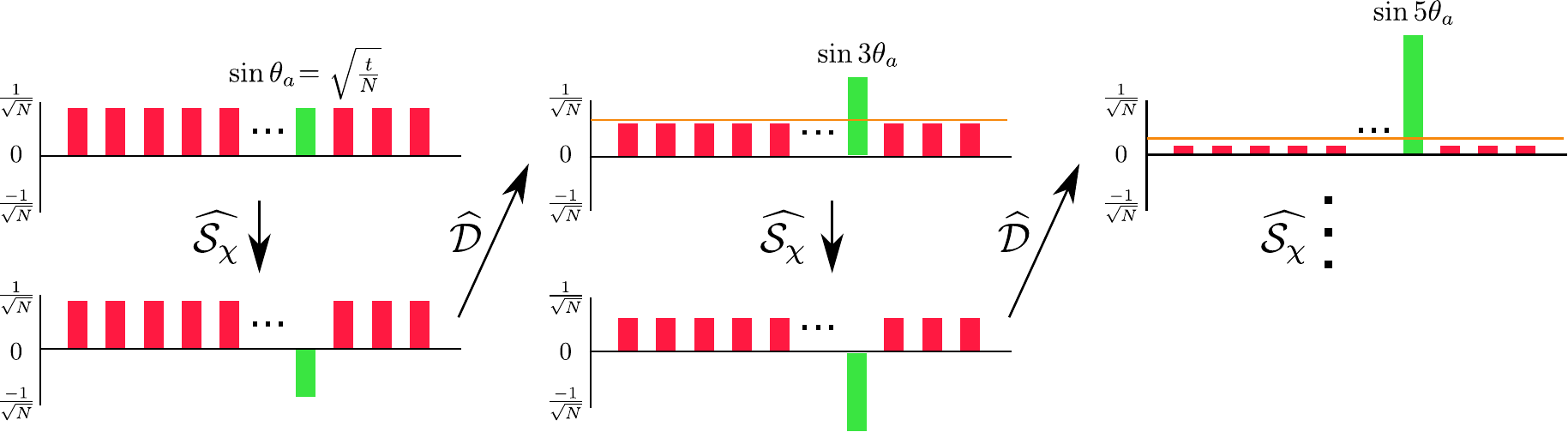}
\caption{Grover's algorithm for $t=1$ and $t=2$. Successive applications of $\widehat{\mathcal{Q}}$ progressively maximize the probability of reading a "good" state, marked in green. The operator $\widehat{\mathcal{S_{\chi}}}$ marks "good" states by flipping their sign, and the operator $\widehat{\mathcal{D}}$ reflects all states around the average, shown as an orange line.}
\label{fig:Grovers}
\end{figure*}

\subsection{Grover's algorithm}
\label{subsec:GroversAlg}

Let the vector $|\Psi\rangle = \widehat{\mathcal{A}}|0\rangle^{\otimes n}|0\rangle$ be expressed as a linear combination of the orthogonal vectors $|\Psi_0\rangle$ and $|\Psi_1\rangle$.  $|\Psi_0\rangle$ contains all the basis states $|i\rangle$ that do not satisfy $f()$, and $|\Psi_1\rangle$ contains the remaining basis states, i.e, those satisfying $f(i)=1$. Clearly, $|\Psi\rangle = \sqrt{\frac{N-t}{N}} |\Psi_0\rangle + \sqrt{\frac{t}{N}}|\Psi_1\rangle$, up to some normalizing constant. More formally this partitions the Hilbert space of the quantum system into two subspaces which are labelled as "good" and "bad" subspaces \cite{Brassard2002}. 

The angle subtended by $|\Psi\rangle$ and $|\Psi_0\rangle$ is denoted by $\theta_a$, with $\sin(\theta_a)= \sqrt{t/N}$ meaning that $|\Psi\rangle$ can also be represented as $|\Psi\rangle = \cos(\theta_a)|\Psi_0\rangle + \sin(\theta_a)|\Psi_1\rangle$. The probability of measuring a good state, i.e. a state in $|\Psi_1\rangle$, is $a = \sin^{2} \theta_{a} = \frac{t}{N}$. Grover's algorithm seeks to maximize this probability, thereby finding a solution to $f()$.

The oracle $\widehat{\mathcal{A}}$ implements $f()$ by marking states which are solutions to $f(x)=1$. Grover's algorithm requires two additional operators. 

$\widehat{\mathcal{S_{\chi}}}$ flips the sign of marked solutions, i.e, those with the output qubit equal to $|1\rangle$ -- see the bottom left image of figure \ref{fig:Grovers}. 

$\widehat{\mathcal{S_{O}}}$, on the other hand, flips the sign of only the zero state coefficient. The diffusion operator $\widehat{\mathcal{D}}=-\widehat{\mathcal{A}}\widehat{\mathcal{S_{O}}}\widehat{\mathcal{A}}^{-1}$ performs a reflection over the mean of all basis states' amplitudes. In practice, this amplifies the amplitude of the marked states, whose amplitude was negative after applying $\widehat{\mathcal{S_{\chi}}}$ -- see the center image in figure \ref{fig:Grovers}, where the orange line indicates the average amplitude of all states and each state is mirrored over this average. Equivalently, this corresponds to rotating state $|\Psi\rangle=\widehat{\mathcal{A}}|0\rangle^{\otimes n}|0\rangle$ by $2\theta_{a}$, thereby increasing the probability of measuring a "good" state from $\sin^{2} (\theta_{a}$) to $\sin^{2} (3\theta_{a})$ -- see the top row of figure \ref{fig:Grovers}.

A Grover evaluation is defined as marking and amplifying the amplitude of "good" states, resulting in Grover's operator $\widehat{\mathcal{Q}} = \widehat{\mathcal{D}}\widehat{\mathcal{S_{\chi}}}$. Figure \ref{fig:Grovers} shows this pictorially, where "bad" states are shown in red, and "good" states in green. Each successive evaluation of $\widehat{\mathcal{Q}}$ further rotates the state vector by $2\theta_a$. After $r$ applications of $\widehat{\mathcal{Q}}$ the angle between $|\Psi\rangle$ and $|\Psi_0\rangle$ is given by $\theta_a^{(r)} = (2r+1)\theta_a$ and $|\Psi\rangle = \cos((2r+1)\theta_a)|\Psi_0\rangle + \sin((2r+1)\theta_a)|\Psi_1\rangle$. The probability of measuring a "good" state is given by 
\begin{equation}
    p_{\Psi_1}(t,r) = \sin^2((2r+1)\theta_a)\text{, with }\theta_a = \arcsin{\sqrt{\frac{t}{N}}}
    \label{eq:GroverSuccess}
\end{equation} 
Since the aim is to maximize $p_{\Psi_1}(t,r)$, $r$ has to be set such that $\sin^{2}((2r + 1)\theta_{a}) \approx 1$, which can be shown to be \(r_{opt} = \left \lfloor{\frac{\pi}{4 \theta_a}} \right \rfloor \). Given that $\theta_a = \arcsin{\sqrt{\frac{t}{N}}}$, then for a sufficiently large $N$ and $t\ll N$ the angle $\theta_a$ will be very small and can be approximated as $\theta_a \approx \sqrt{\frac{t}{N}}$. The expression for the optimal $r$ becomes \[r_{opt} = \left \lfloor{\frac{\pi}{4}\sqrt{\frac{N}{t}}} \right \rfloor \]

This result gives quantum searching the complexity of $\mathcal{O}(\sqrt{\frac{N}{t}})$, a quadratic advantage over the classical complexity of $\mathcal{O}(\frac{N}{t+1})$. Since the $\sin$ is a periodic function, performing more than $r_{opt}$ Grover evaluations will decrease $p_{\Psi_1}$; this periodic behaviour continues with $r$. 

If $t$ is not significantly smaller than $N$, the expression for $r_{opt}$ no longer applies. However, in such cases measuring the uniform superposition $|\Psi\rangle = \widehat{\mathcal{A}}|0\rangle^{\otimes n}|0\rangle$ will succeed with probability $t/N$, which might be computationally effective given that $t/N$ will be significantly larger than 0. By repeatedly sampling the superposition $k$ times, a good state will be found with probability equal to $\frac{t}{N}\sum_{i=1}^{k} \left ( \frac{N-t}{N} \right ) ^{i-1}$. If $t > N/2$, i.e. over half the states are solutions, Grover's algorithm no longer amplifies the amplitude of good states, but a solution can be found by sampling $|\Psi\rangle$ at most twice.


However, the successful application of Grover's algorithm requires that the number of solutions $t$ is known before hand.

\subsection{Adaptive Exponential Search}
\label{subsec:AdaptiveSearchAlg}

When there is an unknown number of solutions $t$, the parameters derived from it, such as $a$, $\theta_a$ and $r_{opt}$, are also unknown. This is the case in many graphics problems; for example, a ray may intersect with an unknown number of primitives. Boyer et  al. \cite{Boyer1998} proposed a hybrid quantum-classical algorithm to this problem, which still allows for a quadratic query complexity gain w.r.t. a classical approach. 

This algorithm is referred to as Adaptive Exponential Search (or \QSearch -- see algorithm \ref{alg:QSearch}) and is based on the classical exponential search algorithm \cite{BENTLEY1976}. It is an iterative algorithm, which executes Grover's algorithm multiple times with different numbers $r$ of evaluations of $\widehat{\mathcal{Q}}$ -- line \ref{algline:QSearchQ} in algorithm \ref{alg:QSearch}. For each iteration $l$ the respective $r_l$ is randomly selected within an interval ranging from $1$ to $M_l$ -- line \ref{algline:QSearch_r_l}. It is this upper limit, denoted by $M_l = \left \lceil{c^l} \right \rceil$, that grows exponentially at each iteration $l$ of \QSearch up to a maximum of $\sqrt{N}$ -- line \ref{algline:QSearch_M_l}. $c$ is a constant such that $1 < c <2$. After each execution of Grover's algorithm, with the respective $r_l$, the resulting state is measured -- line \ref{algline:QSearchMeasure}. The returned value $i$ is then classically evaluated, to check whether $f(i)=1$ (line \ref{algline:QSearch_f_i2}). If the test succeeds the algorithm terminates. This classical evaluation of $f()$ for a single point of its domain is $\mathcal{O}(1)$, therefore it does not add up to the algorithm's complexity. 

Prior to the iterative process the uniform superposition is randomly sampled -- line \ref{algline:QSearchRand}. The measured value is then classically checked; this is the evaluation of $f(i)$ in line \ref{algline:QSearch_f_i1}. Randomly sampling will, with high probability, successfully handle the cases where the number of solutions $t$ is large compared to $N$, particularly the case where $t \geq N/2$. The iterative exponential searching process is entered only when random sampling fails.

This algorithm converges in $\mathcal{O}(\sqrt{\frac{N}{t}})$ steps, therefore it exhibits the same complexity as Grover's algorithm.

\begin{algorithm}[h]
\SetAlgoLined
\KwData{$\widehat{\mathcal{A}}$, $f()$}
$i = $ measure ($\widehat{\mathcal{H}} \ket{0}$) \tcc*{sample the superposition} \label{algline:QSearchRand} 
\eIf{$f(i)==1$}{ \label{algline:QSearch_f_i1}
  found = True \;
}
{ 
  $M_0=0$; $l=0$; found =  False\;
  $c$ constant, such that $1 < c < 2$ \;
  \While{not \text{found} and $M_l<\left \lceil \sqrt{N} \right \rceil$}{
    $l = l+1$\;
    $M_l = \min(\left \lceil{c^l} \right \rceil , \left \lceil{\sqrt{N}} \right \rceil)$ \; \label{algline:QSearch_M_l} 
    $r_l=$ randInt($1 \ldots M_l$)\; \label{algline:QSearch_r_l}
    $\ket{\Psi}=\widehat{\mathcal{Q}}^{r_l} \widehat{\mathcal{A}} \ket{0}$\; \label{algline:QSearchQ}
    $i = $ measure ($\ket{\Psi}$)\; \label{algline:QSearchMeasure}
    \lIf{$f(i)==1$}{ \label{algline:QSearch_f_i2}      found = True    }
  }
}
\Return $i$, found
\caption{QSearch ($\widehat{\mathcal{A}}$, $f(\cdot)$)}
\label{alg:QSearch}
\end{algorithm}

\QSearch never returns a false positive, since solutions are always classically checked at either lines \ref{algline:QSearch_f_i1} or \ref{algline:QSearch_f_i2}. If found=False then it is either a true or a false negative. A true negative corresponds to the no solutions case, i.e., $t=0$. A false negative is understood as returning a value $i$ in the search domain with $f(i)=0$, when in fact $t>0$, i.e., there are solutions $j \neq i$, such that $f(j)=1$. Determining whether $i$ is a false or a true negative cannot be done without classically visiting all points of the search domain (in the worst case). The probability of false negatives, $\overline{p}_{QS}$, can be arbitrarily reduced by repeating \QSearch multiple times. An estimate for $\overline{p}_{QS}$ is developed in section \ref{subsec:Termination}.

\subsection{Minimum finding algorithm}
\label{subsec:MinimumAlg}

The minimum finding algorithm \cite{Durr1996,Ahuja1999} is an iterative approach that searches for a value $i \in X$ which is a solution for $f()$ (equation \ref{eq:solutionIndicator}) and also minimizes a function $g: X \rightarrow \mathbb{Z}$. A new function $f_{min}()$ is defined such that \QSearch can still be used. $f_{min}()$ unites the two conditions imposed by $f()$ and $g()$:
\begin{equation}
  f_{min}(x) =
  \begin{cases}
    1 & \text{if $f(x)==1$ and $g(x) < currMin$} \\
    0 & \text{otherwise.} 
  \end{cases}
  \label{eq:minIndicator}
\end{equation}
$f_{min}()$ returns 1 for all solutions where $g(x) < currMin$ holds, and not only for $g()$'s minimum, which is unknown. 
A new quantum oracle, $\widehat{\mathcal{A}}_{min}$, is defined which marks solutions according to $f_{min}()$: 
\begin{equation}
\widehat{\mathcal{A}}_{min}|0\rangle^{\otimes n}|0\rangle = \frac{1}{\sqrt{N}} \sum_{i=0}^{N-1} |i\rangle |f_{min}(i)\rangle
\label{eq:MinimumOracle}
\end{equation}

\begin{algorithm}[t]
\SetAlgoLined
\KwData{$\widehat{\mathcal{A}}$, $f()$, $g()$, $\#IT$}
$currMin = -1$ \tcc*{no minimum threshold yet}
$solution = -1$; $it = 1$ \;
\While{$it \leq \#IT$}
{
  \eIf{$currMin==-1$}    
  {
    $i$, found = QSearch ($\widehat{\mathcal{A}}$, $f()$)\; \label{algline:QSearchNoMin}
  }
  {
    \tcc{improve on current minimum threshold}
    prepare $f_{min}()$ from $f()$ and $currMin$ \;
    prepare $\widehat{\mathcal{A}}_{min}$ from $f_{min}()$  \;
    $i$, found = QSearch ($\widehat{\mathcal{A}}_{min}$, $f_{min}()$) \; \label{algline:QSearchMin}
  }
  \lIf{found}  {    $currMin = g(i)$; $solution = i$ \label{algline:UpdateMin}  }
  $it += 1$ \;
}
\Return solution
\caption{Minimum finding algorithm}
\label{alg:Minimum}
\end{algorithm}
The minimum finding approach (algorithm \ref{alg:Minimum}) iteratively calls \QSearch in an attempt to find a value less than the current minimum. The actual minimum is unknown, therefore the number of iterations is limited to some constant $\#IT$. Since initially no estimate is known for the minimum, \QSearch is used with operator $\widehat{\mathcal{A}}$ -- line \ref{algline:QSearchNoMin}. Once a solution is found, it is used as the current minimum and operator $\widehat{\mathcal{A}}_{min}$ is used from that point on -- line \ref{algline:QSearchMin}. Whenever \QSearch succeeds in finding a new minimum, which is less that the current threshold, this threshold is updated -- line \ref{algline:UpdateMin}.

This algorithm has been demonstrated to find the minimum, with high probability, with complexity $\mathcal{O}(\sqrt{N})$, therefore maintaining the quadratic query complexity advantage with respect to a classical approach \cite{Durr1996,Ahuja1999}.


%% file: RelatedWork.tex
\section{Related Work}
\label{sec:RelatedWork}

The idea of utilizing quantum mechanics for computation extends back to work by Benioff \cite{benioff1980computer}, Feynmann \cite{Feynman1982} and Deutsch \cite{deutsch1985quantum}. These concepts were built on, and well known algorithms such as solution finding \cite{deutsch1992rapid}, \cite{Grover1996}, and factoring primes \cite{Shor1997} were developed over the subsequent decades.

Closer to the graphics field, a theoretical analysis of minimum finding utilising Grover's and a related method \cite{Durr1996} was performed by \cite{Sadakane2002}. This investigated several geometric applications, such as intersection, nearest neighbour queries, convex hull computation, separation and optimisation problems.

The idea of applying concepts from quantum computing to computer graphics was discussed by Glassner \cite{Glassner2001a,Glassner2001b,Glassner2001c}. Two suggested applications were to accelerate the performance of the Z-Buffer algorithm and ray casting. For both of these problems a similar set up was proposed: a superposition over the primitives in the scene could be created, simultaneous intersections computed, and a minimum could be searched through the application of Grover's algorithm. This would lead to a quadratic advantage in complexity versus a classical linear search of the primitives.

\cite{Lanzagorta2009} also proposed the application of quantum computers to accelerate visibility determining problems, nearest neighbour queries, object-object intersection, radiosity and level of detail. These concepts were further explored by \cite{Caraiman2012} which also proposed using quantum computers to accelerate the Z-Buffer algorithm, ray casting and nearest neighbour queries for photon mapping \cite{jensen1996global}.

These methods make the argument that given certain conditions, such as a large number of primitives $N$ and sufficiently high dimension of the domain $d>=3$, then quantum search methods are asymptotically advantageous over a classical search. In the case of unsorted search, classical computers require $\mathcal{O}(N)$ time and space complexity, whereas quantum algorithms require $\mathcal{O}(\sqrt{N})$ time complexity and $\mathcal{O}(N)$ space complexity. Spatial data structures can gain an improved query time complexity $\mathcal{O}(\log_{d}{N})$, yet require $\mathcal{O}(N\log_{d}{N})$ time to build. As $N$ or $d$ become very large, the building cost can be significant which can decrease the gains from using these structures.

A practical implementation of raycasting for occlusion and visibility (primary rays only) was presented in \cite{alves2019quantum}. This used an orthographic camera and rectangular primitives parallel to the image plane. These algorithms were implemented on a simulator and a simple case was run on a IBM 20 qubit quantum computer. This showed improvements compared to a classical approach, and also illustrated the probabilistic nature of quantum computing in that some visibility and occlusion measurements were incorrect.

While the previous references suggest possible applications of accelerating rendering using quantum computers, none of these, with the exception of \cite{alves2019quantum}, present any practical algorithms or implementations. Our paper is the first to propose a concrete approach and implementation to quantum ray tracing and to present results including direct and indirect lighting, specular reflections and area light sources.

Besides visibility, quantum numerical integration applied to graphics has been addressed by some authors.

Quantum supersampling was investigated by \cite{Johnston2016} who proposed a hybrid quantum-classical method for integrating sub-pixel samples to a final pixel value. This used quantum amplitude estimation combined with a classical lookup table to show improvements compared to classical Monte Carlo integration. This was validated on a simulator, a IBM five qubit machine, and a photonic quantum computer. This work is further elaborated in \cite{Johnston2019}, which also serves as an excellent introductory text for quantum computing.

A different integration approach was applied to the quantum supersampling problem by \cite{shimada2019quantum} who implemented QCoin \cite{abrams1999fast}, another approach for computing integrals, and gained improved results on both a simulator and a real machine. The authors also proposed that the QCoin algorithm potentially could be applied to the simulation of light transport. Quantum numerical integration is a different avenue to the approach proposed in this paper and is likely to require significantly more computation resources as existing algorithms require the entire rendering context including all scene data to reside on the quantum device.


%% file: QRT-Operator.tex
\section{Quantum Ray Tracing: the algorithm}
\label{sec:Raytracing}

This section introduces our approach to implementing a practical quantum ray tracing algorithm. In order to be able to implement this on current hardware and simulators, we need to constrain the complexity of the scenes and the precision of coordinates used. Therefore, we start by describing the geometric setup and processes required for a practical prototype implementation, then describe an intersection operator associated with the scene geometry, and finally put this together into a rendering algorithm.

\subsection{Geometric Setup}
\label{subsec:GSetup}


Existing quantum systems impose severe limits on both the depth and width of executable circuits, i.e., on the number of gates along the circuit's longest path and on the number of qubits, respectively. Additionally, no functional units are included to support arithmetic operations over any numerical data type, such as integer or floating point number representations. Support for these operations must be explicitly included in the user's program, increasing the circuit's depth and width. To maintain circuit's complexity within the limits supported by current simulators, this work imposes the following conditions on the geometric setup:
\begin{enumerate}
    \item all coordinates are positive integers;
    \item geometric primitives are axis aligned rectangles.  
\end{enumerate}
Rays are parameterized by origin, direction and clipping lengths (near and far). The ray's 3D direction vector is the only parameter requiring floating point values; the near and far clipping distances are also integers.
Evaluating a ray/primitive intersection entails two fundamental steps, as described next:
\begin{description}
    \item[eval-rpc()] returns the intersection point between the ray and the plane containing the geometric primitive; this point is evaluated using floating point representation, but the coordinates are then cast to integers -- this point is referred to as {\bf rpc} in the remainder of the manuscript, standing for {\bf r}ay  {\bf p}lane {\bf c}oordinates;
    \item[verify-rpc()] verifies whether the integer coordinates of a ray-plane intersection point (rpc) are contained within the geometric primitive boundaries and within the ray's near and far clipping lengths. 
\end{description}


\subsection{The Intersection Operator}
\label{subsec:IntOperator}

\begin{figure*}
    \centering
    \includegraphics[width=0.75\textwidth]{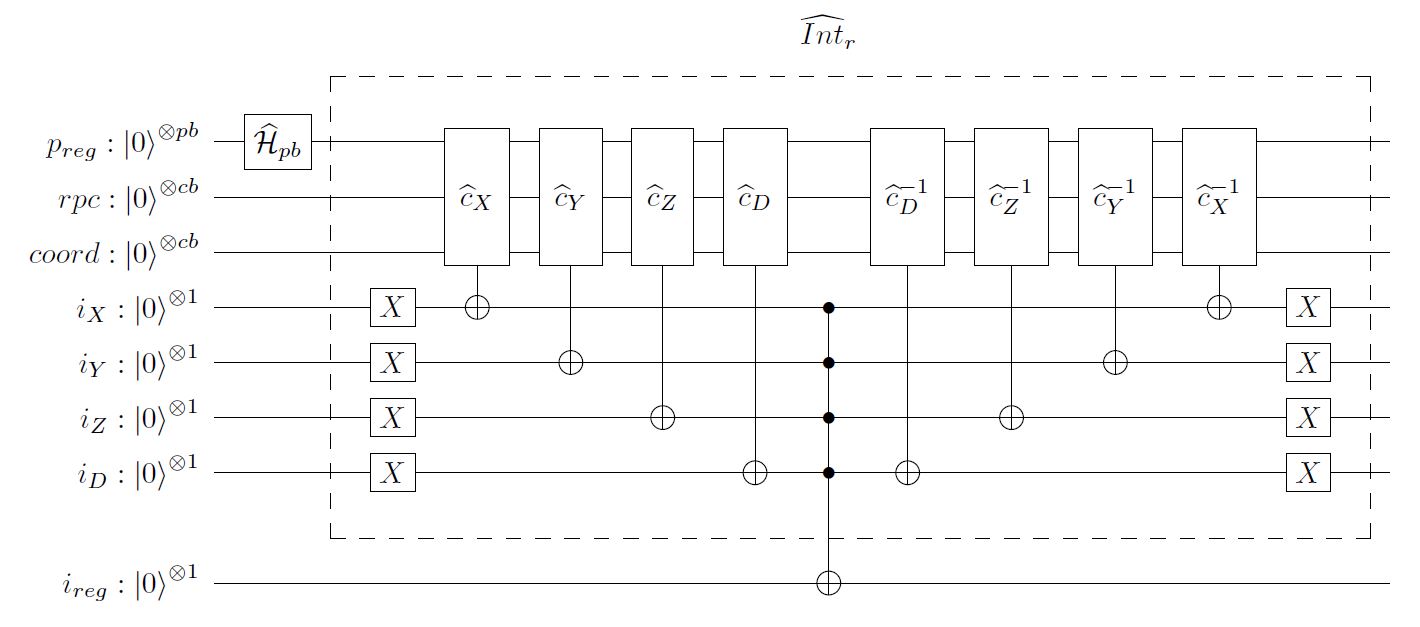}
    \caption{Operator {$\widehat{R}_r$}.}
    \label{fig:Roperator}
\end{figure*}

The ray tracing intersection operator $\widehat{R}_r$ evaluates which primitives, among a set of $N$ primitives, $N= 2^{pb}, \, pb \in \mathbb{N}$, are intersected by ray $r$. It takes as input two quantum registers initialized to $|0\rangle$: $p_{reg}$ with $pb$ qubits and $i_{reg}$ with $1$ qubit. $p_{reg}$ is initially prepared as an uniform superposition, therefore indexing all $N$ primitives. The qubit in $i_{reg}$ is then set to $|1\rangle$ if the corresponding primitive $p$ in the superposition is intersected by $r$:
\begin{equation}
    \widehat{R}_r \underbrace{|0\rangle ^{\otimes pb}}_{p_{reg}} \underbrace{|0\rangle}_{i_{reg}} \mapsto \frac{1}{\sqrt{P}} \sum_{p=0}^{P-1} \left [ \underbrace{|p\rangle}_{p_{reg}} ((1-i_r(p))|0\rangle + i_r(p)|1\rangle ) \right ]
    \label{eq:R_operator}
\end{equation}
Figure \ref{fig:Roperator} presents the quantum circuit for $\widehat{R}_r$ and its component $\widehat{Int}_r$. The superposition on $p_{reg}$ is prepared using $pb$ Hadamard gates (denoted by $\widehat{\mathcal{H}}_{pb}$). The operator $\widehat{Int}_r$ implements, for all primitives $p$ indexed by $p_{reg}$,  the intersection function $i_r(p)$ given by:
\begin{equation}
i_r(p) =
\left\{
	\begin{array}{ll}
		1 & \mbox{if $r$ intersects the primitive with ID $p$} \\
		0 & \mbox{otherwise } 
	\end{array}
    \label{eq:i_function}
\right.
\end{equation}
and \(\widehat{R}_r = \widehat{Int}_r (\widehat{\mathcal{H}}_{pb} \otimes \widehat{\mathcal{I}}_1) \), where $\widehat{\mathcal{I}}_1$ denotes the identity operator applied to the single qubit in $i_{reg}$. 

The operator $\widehat{Int}_r$ is developed by resorting to standard Boolean logic simplification rules and circuit synthesis techniques, while using reversible quantum gates. Due to quantum parallelism an execution of $\widehat{Int}_r$ actually evaluates the intersection of ray $r$ with all $N$ primitives indexed by the superposition in $p_{reg}$. Axes $X$, $Y$, $Z$ and clipping lengths are processed in sequence, as illustrated by the sub circuits $\widehat{c}_X$, $\widehat{c}_Y$, $\widehat{c}_Z$ and $\widehat{c}_D$ in figure \ref{fig:Roperator}. 
\begin{figure*}[htb]
  \centering
    \includegraphics[width=0.75\textwidth]{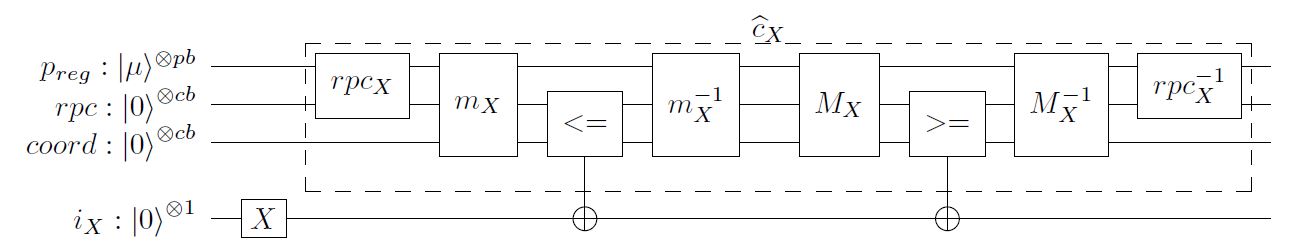}
    \caption{Operator {$\widehat{c}_X$}.}
	\label{fig:cXoperator}
\end{figure*}
Figure \ref{fig:cXoperator} depicts the quantum sub circuit processing axis $X$ and generalizes to $\widehat{c}_Y$ and $\widehat{c}_Z$:
\begin{description}
\item[$rpc_X$ -] generates the $X$ coordinate of the intersection point between the ray and the plane containing the geometric primitive -- see {\bf eval-rpc()} and detailed comments below. For each primitive indexed by $p_{reg}$ it outputs the corresponding ray plane intersection point $X$ coordinate in ancillary register $rpc$, which has $cb$ qubits, with $cb$ being the number of bits required to represent the scene's largest integer coordinate;
\item[$m_X$ -] (respectively $M_X$) generates the minimum (respectively maximum) $X$ coordinate of each primitive indexed by $p_{reg}$ -- primitives are axis aligned rectangles. The output is written in ancillary register $coord$. Equivalent $m_Y$, $M_Y$, $m_Z$ and $M_Z$ operators are used by the $\widehat{c}_Y$ and $\widehat{c}_Z$. These operators can be understood as the loading of the scene description, on a per axis basis, into the quantum circuit. Boolean algebra is used to simplify the operators in terms of the number of logical operations (and, consequently, number of gates);
\item[comparators -] these operators implement {\bf verify-rpc()} for the three reference axis. The $\leq$ operator (respectively $\geq$) negates qubit $i_X$ for those primitives in the superposition where $m_X \leq rpc_X$ (respectively $M_X \geq rpc_X$). Since $i_X$ is initially set to $|1\rangle$ (gate X), its final value will only be $|1\rangle$ if both conditions hold, i.e., if $m_X \leq rpc_X \leq M_X$; it is not possible that neither conditions holds, i.e., $(m_X > rpc_X \; \mathsf{AND} \; M_X < rpc_X)$ \; is impossible;
\item[inverse circuits -] all intermediate computations are reversed using circuits $m_X^{-1}$, $M_X^{-1}$ and $rpc_X^{-1}$, such that ancillary registers $rpc$ and $coord$ final state is $|0\rangle$ and only the quantum state of superposition $p_{reg} \otimes i_X$ is changed by this operator. 
\end{description}

\begin{figure}[htb]
  \centering
    \includegraphics[width=0.59\columnwidth]{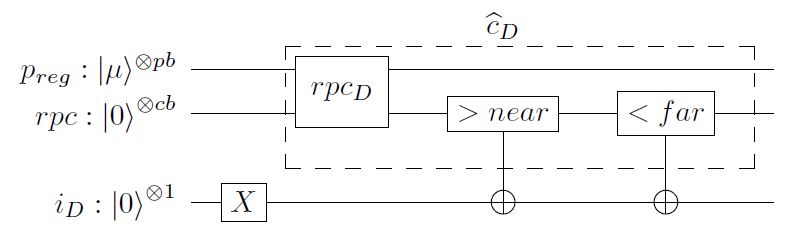}
    \caption{Operator {$\widehat{c}_D$}: $D$ is the ray's direction principal axis ($ D \in \{X, Y, Z\}$).}
	\label{fig:cDoperator}
\end{figure}

Figure \ref{fig:cDoperator} depicts the sub circuit testing whether the ray plane intersection point (for each primitive) is within the ray's near and far clipping distances. These distances are measured along the ray's direction principal axis $D$, which is one of the  world Cartesian coordinate system axis, i.e., $D \in \{X, Y, Z\}$. Sub circuit $rpc_D$ generates the ray plane intersection point coordinate along axis $D$, which is then tested against the ray's near and far parameters. The output is stored in qubit $i_D$. 

Operator $\widehat{Int}_r$ sets qubit $i_{reg}$ to $|1\rangle$ for those primitives in the superposition where {\bf verify-rpc()} holds, i.e., primitives intersected by $r$ with $i_X = i_Y = i_Z = i_D = |1\rangle$. Finally, all computations on $i_X$, $i_Y$, $i_Z$ and $i_D$ are reversed, such that operator $\widehat{R}_r$ only changes the quantum state of  $p_{reg}$ and $i_{reg}$, as described by equation \ref{eq:R_operator}.

{\bf Evaluating ray plane intersection point --} Evaluating {\bf rpc}, the integer coordinates of the intersection point between the ray and the plane containing each primitive (see section \ref{subsec:GSetup}), requires performing a significant number of floating point operations, which is well above current quantum systems' capabilities. There is no theoretical reason why these computations can not be performed by the quantum computer; in fact they can and there are quantum floating point libraries available \cite{Haener2018}. However, including these computations in the quantum circuit would increase its depth, width and number of gates far beyond what can currently be executed in a robust manner. Even simulating such quantum circuits would be compromised due to exponential growth of time and memory requirements. The hybrid quantum classical renderer used for the experiments in this work classically computes these coordinates. The $rpc_X$, $rpc_Y$ and $rpc_Z$ components appropriately set the qubits in $rpc$, but the the coordinates are looked up on a classically evaluated table, rather than evaluated by the quantum circuit. Classically evaluating these coordinates is $\mathcal{O}(N)$, where $N$ is the number of primitives. Quantum evaluation of these coordinates would be $\mathcal{O}(1)$ due to quantum parallelism. Whereas this option seems to compromise this paper's goal of presenting a $\mathcal{O}(\sqrt{N})$ query complexity intersection algorithm, we argue that:
\begin{itemize}
    \item intersection evaluation (verify-rpc()) is still performed using the quantum circuit, therefore exhibiting $\mathcal{O}(\sqrt{N})$ time complexity;
    \item there is no fundamental reason why in the near future these computations can not be performed by the quantum system. As the technology of quantum computers evolves so will the supported circuit depth and width, allowing the migration of these operations to the quantum machine.
\end{itemize}

\subsection{Quantum Rendering Algorithms}
\label{subsec:renderAlg}

\begin{algorithm}[h]
\SetAlgoLined
\KwData{ray;  S -- scene; D -- max ray length}
\lIf{D==MAX\_INT}{DRange=(ray.near, ray.far)}
\lElse{DRange=(ray.near, D)}
$\widehat{R}_r \leftarrow$ build\_R (ray, S, DRange) \;
p, found $\leftarrow$ QSearch ($\widehat{R}_r$, $i_r()$)  \tcp*{$i_r(p)$: -- equation \ref{eq:i_function}} 
rt\_info, depth = eval\_rt\_info (p, ray, S) \;
\Return found, rt\_info, depth
\caption{QTrace(ray, S, D): Tracing a ray}
\label{alg:QTrace}
\end{algorithm}

\QTrace (algorithm \ref{alg:QTrace}) is the method responsible for tracing a ray against the scene's primitives, by interfacing with Grover's adaptive exponential search (\QSearch, algorithm \ref{alg:QSearch}, section \ref{subsec:AdaptiveSearchAlg}).  

\QTrace sets the depth range (\settAlg{DRange}) for the allowed intersections. Depth is hereby understood as the distance from the ray's origin to the intersection point. Ray tracing will strive to find the primitive which is intersected nearest to the ray's origin, i.e., with minimum depth. \settAlg{DRange} is set according to the depth parameter (\settAlg{D}) and to the ray's near and far fields. If \settAlg{D} is equal to some constant \settAlg{MAX\_INT}, then \settAlg{Drange} = $[ray.near \ldots ray.far]$, otherwise \settAlg{Drange} = $[ray.near \ldots D]$. \QTrace then compiles $\widehat{R}_r$ operator's quantum circuit (section \ref{subsec:IntOperator}, figure \ref{fig:Roperator}), based on the current ray, the scene and the depth range. Subsequently, \QSearch is called, with $\widehat{R}_r$ as the oracle, to search for an intersection within the allowed depth range. Upon termination, \QTrace returns whether or not a valid intersected primitive was found together with additional information on this intersection (e.g., primitive ID, 3D point, normal and depth -- all integers).

Shadow rays do not require depth minimization. To evaluate a light source visibility it is enough to assess whether any geometric primitive is intersected within the shadow ray $[\text{near} \ldots \text{far}]$ range. The \QOccluded method is identical to \QTrace without ever reducing the allowed depth range: 

\begin{center}
\settAlg{QOcludded (ray, S) = QTrace (ray, S, MAX\_INT)}. 
\end{center}


\begin{center}
\begin{algorithm}[t]
\SetAlgoLined
\KwData{pixels; rays; S -- scene}
\KwData{\#IT -- maximum iterations; rt\_maps -- rendering data}
D\_map[ALL\_PIXELS] $\leftarrow$ MAX\_INT \;
\For{it $\leftarrow $ 0; it < \#IT; it = it + 1}{
  \ForEach{pix $\in$ pixels}{
    ray $\leftarrow $ rays[pix] \;
    int, rt\_info, depth = QTrace (ray, S, D\_map[pix]) \;  \label{algline:QTrace}
    \lIf{int}{ rt\_maps[pix] = rt\_info; D\_map[pix] = depth; } \label{algline:intFound}
  }
  \tcp{Section \ref{subsec:NN}, alg. \ref{alg:TracePassNN_func} optimization enters here} \label{algline:OptFunc}
}
\Return rt\_maps
\caption{TracePass(pixels, rays, S, \#IT, rt\_maps)}
\label{alg:TracePass}
\end{algorithm}
\end{center}

Algorithm \ref{alg:TracePass}, \TracePass, iteratively handles primary and specular rays (in separate passes). It is an adaptation of the minimum finding algorithm (section \ref{subsec:MinimumAlg}, algorithm \ref{alg:Minimum}), where the quantity being minimized is depth. The algorithm strives to find the primitive which is intersected nearest to each ray's origin. 

It receives as input a list of rays and the list of pixels the rays map to. For the primary rays pass there is one primary ray for each image plane pixel. If specular rays are being processed, then there is a specular ray only for those pixels a specularly reflecting primitive projects into. \QTrace is then used, for each ray, to search for a primitive intersection whose depth is below the current depth threshold (initially set to \settAlg{MAX\_INT}) -- line \ref{algline:QTrace}. If such an intersection is found (line \ref{algline:intFound}) then the current depth threshold is updated and relevant intersection data is stored. The algorithm iterates a user defined number of iterations, \settAlg{\#IT}, in order to allow for convergence to the unknown minimum depth - see section \ref{subsec:MinimumAlg}.   The oracle $\widehat{R}_r$ only accepts intersection depths smaller than the current depth range upper limit. Therefore, even though \QTrace is repeatedly called from \TracePass, it will never return the same intersected primitive twice, since the depth upper limit is constantly updated. After a number of iterations \TracePass will necessarily converge to a state where no further intersections are detected, since either the minimum has been found or no intersection exists.  

Direct illumination evaluation amounts to computing visibility between light source points and the 3D points projecting onto each pixel, without imposing any requirements on depth minimality. The \DirectPass algorithm iteratively calls \QOccluded. If it returns an intersection then the light source is occluded from the current 3D point. But \QOccluded might, with low probability, report no intersection when in fact there are intersections. Repeatedly iterating over \QOccluded diminishes the probability of false negatives. Empirical data has shown that this probability becomes not significant after 2 iterations. \settAlg{DirectPass()}'s detailed algorithm is included in appendix \ref{sec:QRT-algs}, algorithm \ref{alg:DirectPass}.
 
The \RenderScene method (algorithm \ref{alg:RenderScene}, appendix \ref{sec:QRT-algs}) orchestrates all above methods into an hybrid renderer. Rays' trees are traversed breadth first. All primary rays are traced first, followed by eventual specularly reflected rays. Subsequently, light sources' visibility is evaluated and finally pixels shaded accordingly.

%% file: QRT-Results.tex
\section{Quantum Ray Tracing: Analysis}


This section analyses our approach using the operator and algorithms described in section \ref{sec:Raytracing}. Performance scalability with the number of geometric primitives and rendering error with the number of iterations are analysed and compared to the classical approach.

\subsection{Experimental Setup}
\label{sec:experimental_setup}

All experiments were performed using Qiskit (ver. 0.18.1), an open source SDK for quantum computing at the level of algorithms, circuits and pulses. Qiskit allows executing the quantum circuits both on a real quantum backend or on a simulator. All results presented in this paper were obtained using Qiskit's simulator, since the number of gates and the depth of the developed circuits are well above the capabilities of current real quantum computers (this is demonstrated in section \ref{subsec:RealDevice}). In order to maintain the number of required qubits within simulators' supported ranges ($\leq 25)$ the world space is restricted to a 16x16x16 integer coordinates volume; this quantization of the 3D world coordinates is clearly perceivable in all rendered images and severely constrains the complexity of the 3D scenes in terms of the number and relative positioning of geometric primitives. Additionally, simulation times are huge and directly proportional to the number of rays; all images were rendered at 128x128 resolution to guarantee acceptable (a few hours) rendering times. 

\begin{figure}[ht]
\begin{center}
	\subfigure[Qornell Box scene.]{
	    \label{fig:DepthMap-Qornell}
        \includegraphics[width=0.3945\columnwidth]{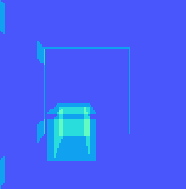}
    } 
	\subfigure[Depth Complexity scene.]{
	    \label{fig:DepthMap-DComp}
        \includegraphics[width=0.4\columnwidth]{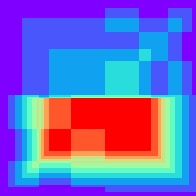}
    } 
	\subfigure{
	    \label{fig:DepthMap-Colorbar}
        \includegraphics[width=0.07\columnwidth]{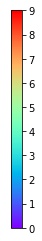}
    }  \\
    \caption{Image space depth complexity maps at 128x128 resolution. This illustrates how many primitives project into each pixel.}
\label{fig:DepthMaps}
\end{center}
\end{figure}

Experimental data sets include the following scenes:
\begin{description}
\item[Quantum Cornell Box - ] denoted by the Qornell Box (figure \ref{fig:C-Qornell064-WH128}). The scene is lit by two point light sources and includes a specular mirror in the left wall. The respective image space depth complexity map (i.e. how many primitives project into each pixel) is presented in figure \ref{fig:DepthMap-Qornell}. Several versions are used, with the number of geometric primitives ranging from 8 to 512. However, for all versions only 14 such primitives are within the viewing frustrum and visible from the camera\footnote{the 8 geometric primitives scene is further simplified by removing back facing polygons of the interior cube and by simplifying the mirrored wall}; the image space depth complexity is therefore independent on the number of primitives;
\item[Depth Complexity - ] this synthetic scene, with all primitives parallel to the image plane, exhibits varying depth complexity (figure \ref{fig:DepthMap-DComp}). The upper left and lower right quadrants have a depth complexity of 2 and 9, respectively, with smooth variations around the borders due to perspective projection. The remaining two quadrants have similar depths, but a checkerboard pattern of polygons is included to increase the image plane's spatial frequency. The impact of different depth complexities across the image plane in the number of \TracePass iterations (and consequently, oracle evaluations) will be assessed in section \ref{subsec:MinimumIT}. For scalability analysis there are 2 versions of the scene, with 32 and 64 primitives, both exhibiting exactly the same depth complexity and rendering to the same image -- figure \ref{fig:Depth-SN-NN}.
\end{description}

\subsection{Performance and Error Metrics}

The expected advantage of hybrid quantum/classical ray tracing over classical ray tracing is a quadratic improvement on the asymptotic number of intersection evaluations as the number of the scene's geometric primitives increases. The following metrics are used to quantify both the classical and the quantum renderers' performance: number of intersection evaluations, number of rays traced (\#Rays) and the respective ratio. The number of intersection evaluations for the classical case is the number of calls to the intersection method (\#C\_Int) and equals the number of rays times the number of geometric primitives (since no acceleration structures are used to impose a spatial ordering). For the quantum case, the number of intersection evaluations (\#Int) is the number of oracle evaluations (\#Eval) plus the number of classical intersection evaluations (\#C\_Int); the latter is required because \QSearch classically verifies the results measured from the quantum circuit (lines \ref{algline:QSearch_f_i1} and \ref{algline:QSearch_f_i2} in algorithm \ref{alg:QSearch}). These metrics are summarized in table \ref{tab:metrics}.   

\begin{table}[ht]
  \caption{Performance and error metrics.}
  \label{tab:metrics}
  \begin{center}
  \begin{tabular}{l|p{0.75\linewidth}}
    \toprule
    \multicolumn{2}{c}{Performance metrics} \\     \midrule
    \#Rays & Number of rays traced \\
    \#C\_Int & Classical: number of calls to the ray/geometry intersection method \\
    \#Eval & Quantum: number of evaluations of the intersection oracle \\
    \#Int & \#C\_Int + \#Evals: total number of intersection evaluations (classical + quantum) \\
    Int/Ray & performance measure \\
    \#It & number of TracePass() iterations \\
    \#Cpix & number of pixels updated by exploiting image space coherence (sec. \ref{subsec:NN}) \\ \midrule
    \multicolumn{2}{c}{Error metrics} \\     \midrule
    \#NRMSE & Normalized Root Mean Squared Error between the classical and the quantum rendered images \\
    \#Dpix & number of pixels with different RGB values \\
    \%Dpix & percentage of pixels with different RGB values \\
\bottomrule
\end{tabular}
\end{center}
\end{table}

Since the classical renderer is fully deterministic, the classically rendered images are used as the reference result for the images obtained with the quantum renderer under the same rendering conditions. To quantify the accuracy of the quantum renderer with respect to the classical, the Normalized Root Mean Square Error (NRMSE) is used. The number of pixels in the quantum image with different RGB values from the classical image is also reported (\#Dpix);  this metric provides further insight into the ability of the quantum renderer to correctly compute visibility along rays' trees. 

\subsection{Scalability with the number of primitives}
\label{subsec:prim_scale}

\begin{table}[t]
  \caption{Scalability with the number of primitives.}
  \label{tab:ScaleNumPrims}
  \begin{minipage}{0.9\columnwidth}
  \begin{center}
  \begin{tabular}{r|r|rr|rr}
    \toprule
  &  & \multicolumn{2}{|c|}{Classical} & \multicolumn{2}{c}{Quantum} \\
 Scene & N & \#Int & $\frac{\text{Int}}{\text{Ray}}$ & \#Int &  $\frac{\text{Int}}{\text{Ray}}$ \\
    \midrule
     \multirow{7}{*}{Qornell}& 8 & 341 K & 8 & 500 K & 12.0  \\
     & 16 & 669 K & 16 & 731 K & 18.0  \\
     & 32 & 1339 K & 32 & 1112 K & 27.4  \\
     & 64 & 2678 K & 64 & 1366 K & 33.6  \\
     & 128 & 5356 K & 128 & 2045 K & 50.4  \\
     & 256 & 10712 K & 256 & 2075 K & 51.3  \\
     & 512 & 21423 K & 512 & 3240 K & 79.7 \\ \midrule
     \multirow{2}{*}{Depth} & 32 & 1207 K & 32 & 1132 K & 30.7 \\
     & 64 & 2413 K & 64 & 1455 K & 39.4 \\
\bottomrule
\end{tabular}
\end{center}
\smallskip
\end{minipage}
\end{table}

\begin{figure}[ht]
\begin{center}
    \includegraphics[width=0.5\columnwidth]{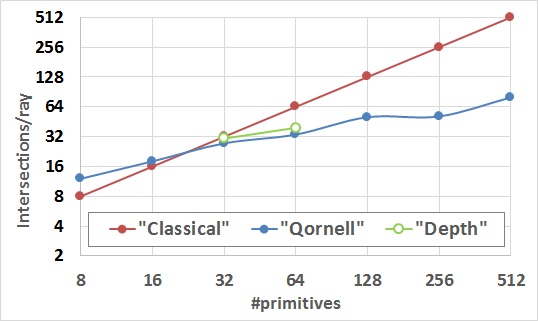}
    \caption{Qornell Box: number of oracle evaluations per ray with varying numbers of geometric primitives}
\label{fig:ScaleNumPrims}
\end{center}
\end{figure}

Table \ref{tab:ScaleNumPrims} presents the average number of intersection evaluations per ray with varying number of geometric primitives. While this figure increases linearly for the classical renderer, it grows with the square root of the number of geometric primitives in the quantum case. It is clearly shown for both scenes (although with only two data points for the Depth scene corresponding to the two versions of the scene described in section \ref{sec:experimental_setup}) that the quantum renderer scales quadratically better than the classical renderer with increasing number of geometric primitives. Figure \ref{fig:ScaleNumPrims} visually illustrates this trend. 

Figure \ref{fig:Q_Qornell064_128x128_MP4} presents the rendered image for the 64 primitives case. This is mostly identical to the reference case (figure \ref{fig:C-Qornell064-WH128}) except for: i) a very reduced number of pixels scattered across the image plane and ii) a significant number of pixels in the intersection between the back wall and the adjacent walls. The former is due to the stochastic nature of the quantum minimum finding algorithm: a false positive (non intersecting primitive) is occasionally measured, despite Grover's amplitude amplification. The ensuing classical verification transforms this measurement into a non-intersection by verifying that the ray does not intersect that primitive -- the resulting false negative appears as an erroneous pixel. The latter is due to Z fighting: the two intersecting walls are intersected by the primary ray at exactly the same (integer) ray length. In the classical case, the visible primitive is deterministically selected following the sequential order in which primitives are stored in the respective data structure. In the quantum case equal depth primitives are stochastically selected with equal probabilities.  

These experimental results corroborate the theoretical claim that the number of intersection evaluations is $\mathcal{O}(\sqrt{N})$, a clear advantage over the $\mathcal{O}(N)$ complexity of the classical approach.

\subsection{Minimum finding: number of iterations}
\label{subsec:MinimumIT}

Table \ref{tab:error_passes} presents a range of experimental statistics and error measures for the Qornell Box and the Depth Complexity scenes, both with 64 geometric primitives. Results are presented for different numbers of  \TracePass iterations. However, only two iterations are performed for shadow rays, since these mostly converge after two trials. 

\begin{table}[t]
  \caption{Statistics and error measures for different number of iterations (both scenes with 64 geometric primitives)}
  \label{tab:error_passes}
  \begin{center}
  \begin{tabular}{cr|rr|rrr}
  \multicolumn{2}{c}{} & \multicolumn{2}{|c|}{} & \multicolumn{3}{|c}{Error Metrics} \\
Scene & It & \#Int & $\frac{\text{Int}}{\text{Ray}}$ & NRMSE & \#Dpix & \%Dpix \\
    \midrule
    \multirow{5}{*}{Qornell} & 1 & 631 K & 17.0 & 63\% &  2278 & 14\% \\
     & 2 & 905 K & 22.6 & 24\% & 430 & 3\%\\
     & 3 & 1139 K & 28.1 & 10\% & 146 & 1\% \\
     & 4 & 1366 K & 33.6 & 4\% & 107 & 1\% \\
     & 5 & 1593 K & 39.2 & 1\% & 103 & 1\% \\ \midrule
     \multirow{6}{*}{Depth} & 1 & 552 K & 15.9 & 34\% & 5665 & 35\% \\
     & 2 & 789 K & 21.7 & 13\% & 2556 & 16\%\\
     & 3 & 1017 K & 27.6 & 4\% & 973 & 6\%\\
     & 4 & 1240 K & 33.6 & 1\% & 304 & 2\%\\
     & 5 & 1455 K & 39.4 & 0\% & 95 & 1\%\\
     & 6 & 1668 K & 45.2 & 0\% & 47 & 0\%\\
     \bottomrule
\end{tabular}
\end{center}
\smallskip
\end{table}

Table \ref{tab:error_passes} data allows for a clear understanding of how many intersection evaluations (quantum plus classical) are required, in average, per ray to render the respective image within a given error bound. In particular, by comparing \#Dpix for both scenes it becomes clear that the number of \TracePass iterations (and, consequently, the number of intersection evaluations per ray) is dependent on the depth complexity. \#Dpix is much higher for the Depth Complexity scene than for the Qornell Box scene, for the same number of iterations, because the former exhibits higher depth complexity. Since the number of primitives intersected by a ray is higher on average, i.e., the number of solutions, $t$, satisfying the intersection operator $\widehat{R}_r$ is larger for the initial iterations, more iterations are required to converge towards the minimum depth solution. 

NRMSE is larger for the Qornell Box scene than the Depth Complexity scene, because the former includes a mirror spawning an additional layer of specular rays. Additionally, \#Dpix does not go below 100 for the Qornell Box (at 128x128 resolution) due to Z fighting (see section \ref{subsec:prim_scale}).

\begin{figure}[htp]
\begin{centering}
	\subfigure[1 iteration (NRMSE=63\%).]{
	    \label{fig:Q_Qornell064_128x128_MP1}
        \includegraphics[width=0.3\columnwidth]{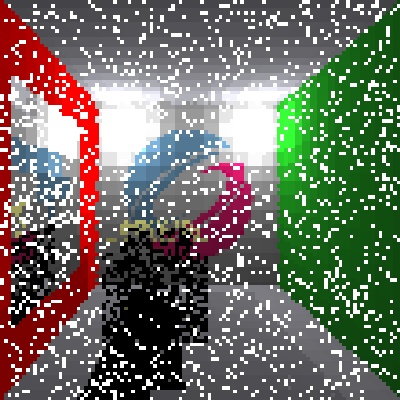}
    } \hspace{0.01cm}
	\subfigure[2 iterations (NRMSE=24\%).]{
	    \label{fig:Q_Qornell064_128x128_MP2}
        \includegraphics[width=0.3\columnwidth]{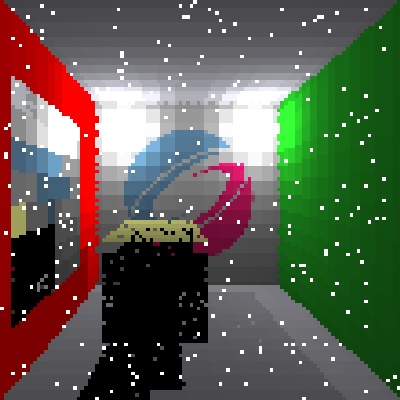}
    } \hspace{0.01cm}
    \subfigure[5 iterations (NRMSE=1\%).]{
	    \label{fig:Q_Qornell064_128x128_MP5}
        \includegraphics[width=0.3\columnwidth]{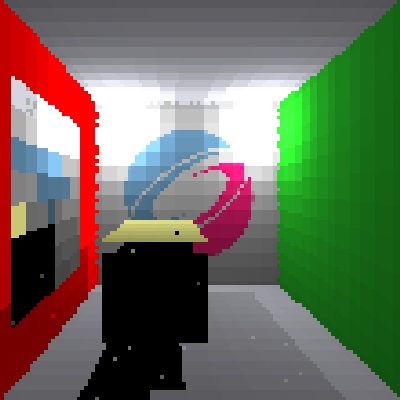}
    } \\
	\subfigure[Difference : 2278 pixels.]{
	    \label{fig:Q_Qornell064_128x128_MP1_Diff}
        \includegraphics[width=0.3\columnwidth]{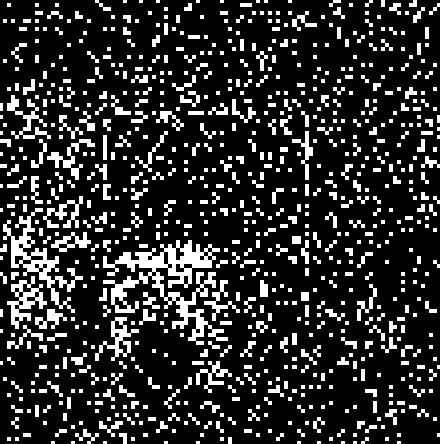}
    } \hspace{0.01cm}
	\subfigure[Difference : 430 pixels.]{
	    \label{fig:Q_Qornell064_128x128_MP2_Diff}
        \includegraphics[width=0.3\columnwidth]{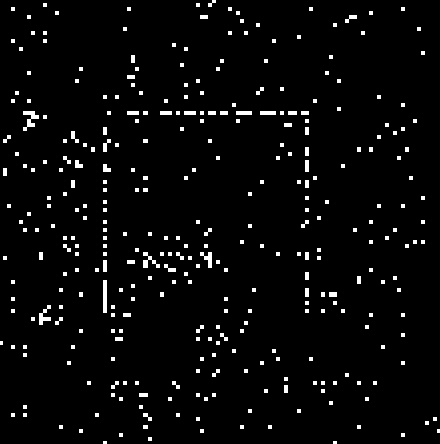}
    } \hspace{0.01cm}
	\subfigure[Difference : 103 pixels.]{
	    \label{fig:Q_Qornell064_128x128_MP5_Diff}
        \includegraphics[width=0.3\columnwidth]{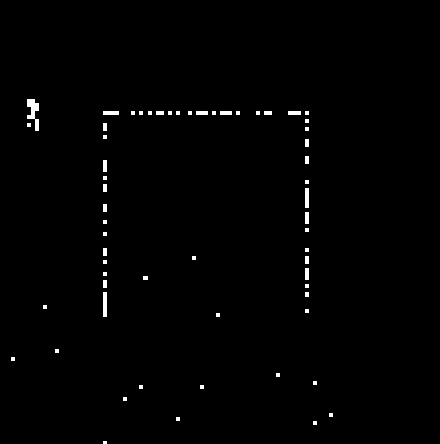}
    }
    \caption{\TracePass number of passes: images and difference to reference - Qornell Box with 64 primitives}
\label{fig:MinimumPassesQornell}
\end{centering}
\end{figure}

\begin{figure}[ht]
\begin{centering}
	\subfigure[1 iteration (NRMSE=34\%).]{
	    \label{fig:Q_Depth064_128x128_MP1}
        \includegraphics[width=0.3\columnwidth]{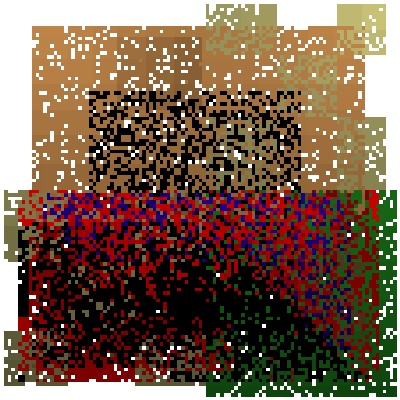}
    } \hspace{0.01cm}
    \subfigure[3 iterations (NRMSE=4\%).]{
	    \label{fig:Q_Depth064_128x128_MP3}
        \includegraphics[width=0.3\columnwidth]{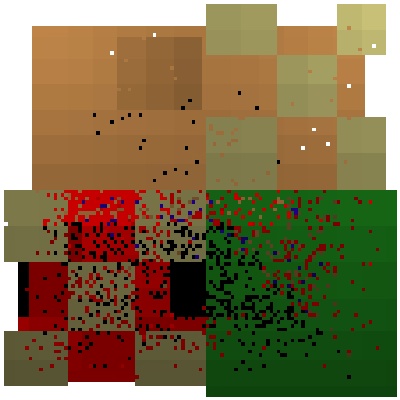}
    } \hspace{0.01cm}
	\subfigure[6 iterations (NRMSE=0\%).]{
	    \label{fig:Q_Depth064_128x128_MP6}
        \includegraphics[width=0.3\columnwidth]{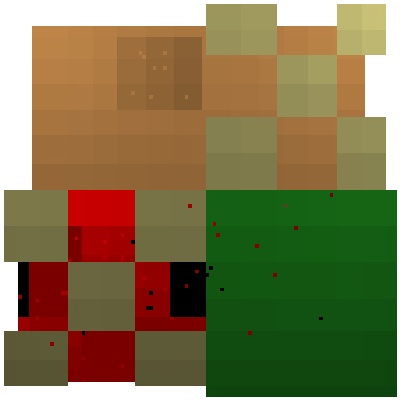}
    } \\
	\subfigure[Difference : 5665 pixels.]{
	    \label{fig:Q_Depth064_128x128_MP1_Diff}
        \includegraphics[width=0.3\columnwidth]{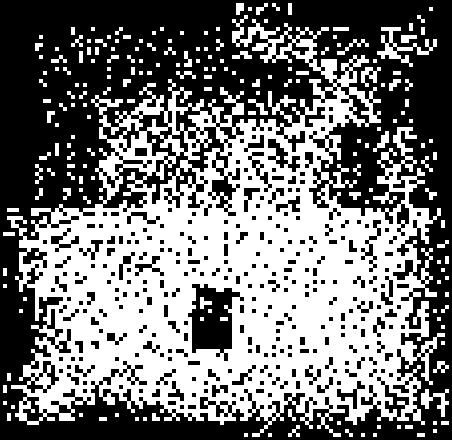}
    } \hspace{0.01cm}
	\subfigure[Difference : 973 pixels.]{
	    \label{fig:Q_Depth064_128x128_MP3_Diff}
        \includegraphics[width=0.3\columnwidth]{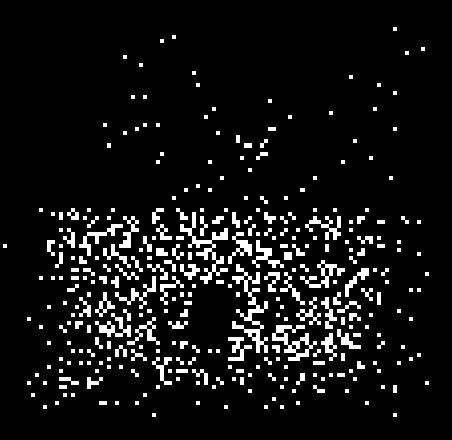}
    } \hspace{0.01cm}
	\subfigure[Difference : 47 pixels.]{
	    \label{fig:Q_Depth064_128x128_MP6_Diff}
        \includegraphics[width=0.3\columnwidth]{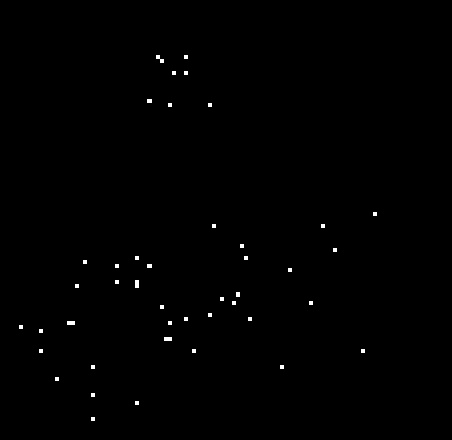}
    }
    \caption{\TracePass number of passes: images and difference to reference - Depth Complexity with 64 primitives}
\label{fig:MinimumPassesDepthComplexity}
\end{centering}
\end{figure}

Figures \ref{fig:MinimumPassesQornell} and \ref{fig:MinimumPassesDepthComplexity} further illustrate that errors concentrate in image regions with larger depth complexity. As more iterations are executed the minimum threshold progresses towards the correct value. The challenge is that it is never known whether the minimum, and therefore the sought solution, has already been found or not.

\begin{table}[b]
  \caption{Individual iteration statistics for the Depth Complexity scene with 6 iterations -- primary rays only}
  \label{tab:error_individual_iteration}
  \begin{center}
  \begin{tabular}{r|rrrrr}
\#Rays & It & \#Int & $\frac{\text{Int}}{\text{Ray}}$ & \%Dpix & Dpix\\
    \midrule
    \multirow{6}{*}{16384} & 1 &  139 K & 8.5 & 35.7\% & 5853 \\
     & 2 & 178 K & 10.8 & 16.1\% & 2630 \\
     & 3 & 199 K & 12.2 & 6.4\% & 1047 \\
     & 4 & 208 K & 12.7 & 1.9\% & 306 \\
     & 5 & 212 K & 12.9 & 0.5\% & 83 \\
     & 6 & 213 K & 13.1 & 0.3\% & 47 \\
     \bottomrule
\end{tabular}
\end{center}
\smallskip
\end{table}

Table \ref{tab:error_individual_iteration} presents statistics for primary rays for each individual iteration out of a total of 6 iterations for the Depth Complexity scene. \%Dpix decreases with the iteration count, as expected. The relevant data, however, is that the number of intersections per ray at each iteration also increases with the iteration count. As primitives closer to the ray's origin are found by early iterations, there will be less primitives intersected by the ray and closer to its origin than the current minimum (i.e., within the allowed depth range, which shrinks at each iteration -- see section \ref{subsec:renderAlg} and algorithm \ref{alg:TracePass}). The number of primitives satisfying the oracle criterion diminishes as more iterations are performed, reaching 0 when the visible primitive is found. Grover's adaptive search algorithm, \QSearch, will keep searching for a non-existent solution, easily reaching the maximum number of allowed oracle evaluations. Table \ref{tab:error_individual_iteration} shows that after the first iteration the correct primitive has not been found for only 5853 pixels; still all 16384 pixels are processed with \QSearch, which will maximize computations for 10531 of these pixels for which no solution exists. This is a major source of wasted computation: as soon as the visible primitive is found, further passes will perform more work searching for a better solution that does not exist.

%% file: QRT-Optimizations.tex
\section{Optimized Hybrid Ray Tracing}
\label{sec:Optimize}

This section improves on previous results by exploit two sources of information arising from the data. Firstly, image plane coherence is exploited to improve convergence onto the minimum depth solution, i.e., the visible geometric primitive at each pixel. Secondly, a principled termination criterion based on successive non intersections for the same ray is proposed, enabling autonomously stopping iterating \TracePass on a per ray basis.
A comparison with a randomized classical renderer with complexity $\mathcal{O}(\sqrt{N})$, i.e., similar to the quantum renderer's complexity, is also presented. Results show that the quantum approach achieves higher image quality for identical computational effort (number of intersection evaluations per ray), due to the amplitude amplification process associated with quantum searching. 

\subsection{Gathering Neighboring Data}
\label{subsec:NN}

Natural images most often exhibit coherence across the image plane, meaning that some measure is locally constant, i.e., varies smoothly within a neighborhood of each pixel \cite{Groelle1995}. Examples of such measures are radiance and the geometric primitives that project onto points pierced by primary rays (pixels, within the context of this paper). This section exploits the latter: given image plane coherence, it is highly probable that the geometric primitive projecting onto any pixel (x,y) also projects onto at least one of its 4 von Neumann neighbours. To benefit from image space coherence the proposed optimization gathers information for each pixel from its immediate neighbours. Function \NeighOpt (algorithm \ref{alg:TracePassNN_func}) is called for each iteration of \TracePass after tracing all rays (algorithm \ref{alg:TracePass}, line \ref{algline:OptFunc}).

\begin{algorithm}[h]
\SetAlgoLined
  \ForEach{pix $\in$ pixels}{
    ray $\leftarrow $ rays[pix] \;
    primID = rt\_maps[pix].primID \;
    \ForEach{neigh\_pix $\in$ pix.neighbours()}{
      neigh\_primID = rt\_maps[neigh\_pix].primID \;
      intersects, depth = classic\_intersect (ray, neigh\_primID) \; \label{algline:neighprim_test}
      \If{intersects AND depth < depth\_map[pix]}{   
        Update all info for pix: rt\_maps, depth\_map[pix] \; \label{algline:neighprim_update}
      }
    }
}
\caption{NeighOpt(pixels, rays, rt\_maps, depth\_map)}
\label{alg:TracePassNN_func}
\end{algorithm}

For both primary and specular rays \NeighOpt classically intersects against each pixel's ray the primitives currently projected onto each of its 4 neighbours  -- line \ref{algline:neighprim_test}. If any such primitives is intersected, then the one at minimum depth is taken as the one projecting onto the current pixel -- line \ref{algline:neighprim_update}. This process allows local information sharing, accelerating convergence towards local minimum depth, therefore reducing the number of \TracePass iterations on a per ray basis. Furthermore, since the update is performed '{\em in place}' a diffusion process occurs, where previously found true minimum depth primitives can spill to contiguous regions. 

\begin{figure}
\begin{center}
	\subfigure[NRMSE = 0.3\%, Int/Ray = 17.4.]{
	    \label{fig:Q_Qornell064_128x128_NN_MP1}
        \includegraphics[width=0.37\columnwidth]{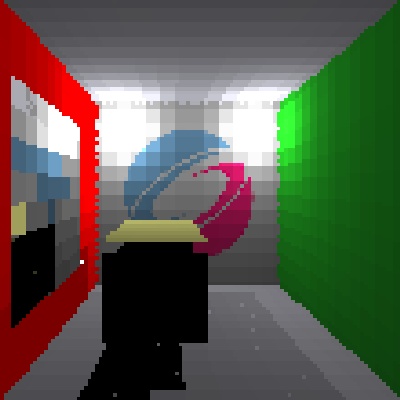}
    } \hspace{0.3cm}
	\subfigure[Errors: \#Dpix = 102; \#Cpix = 2542.]{
	    \label{fig:Q_Qornell064_128x128_NN_MP1_Diff}
        \includegraphics[width=0.37\columnwidth]{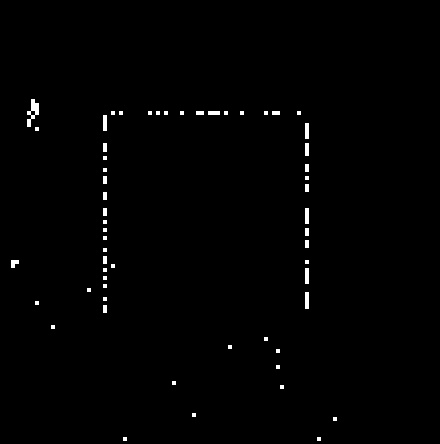}
    } \\
	\subfigure[NRMSE = 0.0\%, Int/Ray= 18.4.]{
	    \label{fig:Q_Depth064_128x128_NN_MP1}
        \includegraphics[width=0.37\columnwidth]{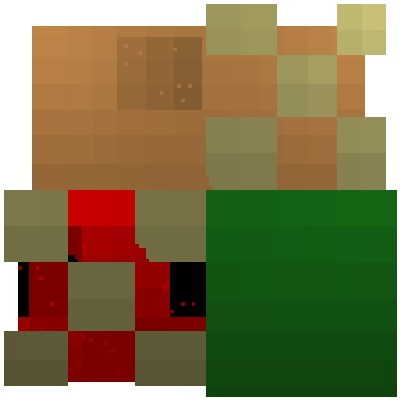}
    } \hspace{0.3cm}
	\subfigure[Errors: \#Dpix = 50; \#Cpix = 5817.]{
	    \label{fig:Q_Depth064_128x128_NN_MP1_Diff}
        \includegraphics[width=0.37\columnwidth]{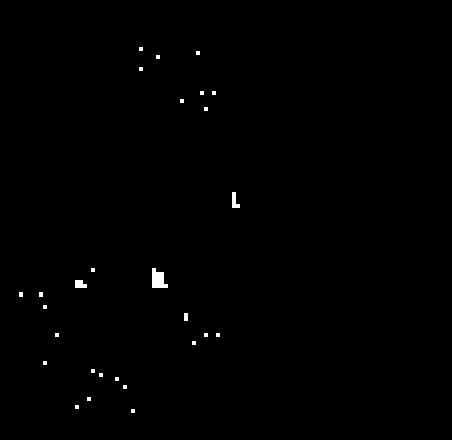}
    } \\
    \caption{Gathering neighbouring data: single iteration}
\label{fig:NNGather}
\end{center}
\end{figure}

Table \ref{tab:error_passesNN} presents the average ratio of intersections per ray for different numbers of optimized \TracePass iterations. It also presents NRMSE, \#Dpix and \#Cpix (the number of pixels updated by the optimization process within each iteration -- see also table \ref{tab:metrics}). By comparing with table \ref{tab:error_passes} it is clear that the optimization process at the end of the first iteration is able to update most pixels to the correct visible geometric primitive. Using the Depth scene as the illustrative example, at the end of the first iteration without optimization, there were in excess of 5500 pixels different from the reference solution; by gathering data from neighbouring pixels around 5800 pixels were updated and only 50 differ from the reference solution. This huge improvement in the convergence rate is further illustrated in figure \ref{fig:NNGather}, where rendered images for both scenes are presented for a single optimized iteration; visual comparison with the first row of figures \ref{fig:MinimumPassesQornell} and \ref{fig:MinimumPassesDepthComplexity} clearly demonstrate the achieved gain.
\begin{table}[b]
  \caption{Statistics and error measures for different number of iterations while gathering neighbouring data (both scenes with 64 geometric primitives)}
  \label{tab:error_passesNN}
  \begin{minipage}{\columnwidth}
  \begin{center}
  \begin{tabular}{c|c|r|rrr}
Scene & It & $\frac{\text{Int}}{\text{Ray}}$ & NRMSE & \#Dpix & \#Cpix \\
    \midrule
    \multirow{3}{*}{Qornell}  &  1 & 17.4 & 0.3\% & 102 & 2542\\
     & 2 & 23.0 & 0,3\% & 96 & 0 \\
     & 3 & 28.7 & 0.3\% & 97 & 0 \\ \midrule
    Depth &  1 & 18.4 & 0.0\% & 50 & 5817\\
    Complexity &  2 & 24.2 & 0.0\% & 29 & 4\\
     & 3 & 29.9 & 0.0\% & 17 & 0 \\
     \bottomrule
\end{tabular}
\end{center}
\smallskip
\end{minipage}
\end{table}

However, there is a subtle performance penalty with this approach: each optimized iteration requires evaluating more classical intersections than in the non-optimized case. For each pixel, the primitives projecting into its 4 immediate neighbours, if different, have to be classically checked for intersection against the current ray. This increase in the rate of intersections per ray can be observed by comparing columns $\frac{\text{Int}}{\text{Ray}}$ of tables \ref{tab:error_passes} and \ref{tab:error_passesNN}. In practice, however, the optimized process requires far fewer iterations to converge to a similar quality result. Therefore, the total number of intersection evaluations (and the respective ratio to the number of rays) is always much less than for the non-optimized approach.

\subsection{Termination Criterion}
\label{subsec:Termination}

The current approach suffers from two major flaws. The number of \TracePass iterations is fixed and set by the user as parameter \#IT. Additionally, each iteration processes all rays (primary or specular, depending on the pass) independently of how likely it is that a solution has already been found for each particular ray. A criterion is required to autonomously, and per ray, terminate iterations. 

Once the geometric primitive closer to the rays' origin has been found, any further call to \QTrace will always return a true negative result (no intersection). However, given the probabilistic nature of quantum search, there is a non-zero probability that \QTrace returns a false negative, i.e., reports no intersection when in fact there are intersected geometric primitives closer to the rays' origin than the current minimum depth threshold. The rational behind the proposed per ray termination criterion is as follows: 
\begin{itemize}
    \item if an intersection is found, update the visible primitive ID and the minimum depth threshold and trace the ray again in the next iteration;
    \item if no intersection is found, then add 1 to the number of consecutive false negatives observed for this ray ($sn_{ray}$). Then draw a random number, $\xi_{ray}$, uniformly distributed in $[0 .. 1)$; if $\xi_{ray} \leq \left ( \overline{p}_{QS} \right )^{sn_{ray}}$, then trace that ray again in the next iteration, otherwise terminate it. $\overline{p}_{QS}$ is an estimate of observing a false negative.
\end{itemize} 
\TracePass will continue iterating for non terminated rays until there is none. 

An estimate of $\overline{p}_{QS}$, i.e., the probability that \QSearch() (algorithm \ref{alg:QSearch}) returns a false negative, is derived next. For any given iteration $l$ of \QSearch, the number of evaluations of Grover's operator $\widehat{\mathcal{Q}}$ is randomly selected with uniform distribution, such that $r_l \in \{1 \ldots M_l\}$. The expected value for $r_l$ is therefore $M_l/2$.  Combining this result with equation \ref{eq:GroverSuccess}, the success probability of iteration $l$ for any particular $t$ is \[p_l(t) = \sin^2((M_l+1)\theta_a)\] The probability of being unsuccessful, i.e., of iteration $l$ reporting a false negative when in fact there are $t$ solutions, is 
\begin{equation}
    \overline{p}_l(t) = 1 - p_l(t) = \cos^2((M_l+1)\theta_a)
    \label{eq:Iteration_FNeg_t}
\end{equation}
$t$ is unknown, but let $p_t(i)$ be the probability mass function, with $p_t(i) = P(t=i), i = 1 \ldots N$.  The probability of a false negative for any possible $t$ is then given by 
\begin{equation}
    \overline{p}_l = \sum_{t=1}^N \left ( p_t(t) \cos^2((M_l+1)\theta_a) \right )
    \label{eq:Iteration_FNeg}
\end{equation}
In order to return a false negative, algorithm \ref{alg:QSearch} must iterate $L$ times, with the iteration index $l$ taking all values $l \in \{1 \ldots L\}$ and $M_l$ taking all corresponding values from set $\mathcal{M} = \{ \left \lceil c^1 \right \rceil,  \left \lceil c^2 \right \rceil,  \ldots, \min(\left \lceil{c^L} \right \rceil , \left \lceil{\sqrt{N}} \right \rceil)\}$. $L$ and $\mathcal{M}$ can be easily computed given $c$ and $N$. The probability that the algorithm returns a false negative after performing all $L$ iterations is therefore given by a variant of the Poisson binomial distribution:

\begin{equation}
    \overline{p}_{QS} = \prod_{M_l \in \mathcal{M}} \overline{p}_l \nonumber 
    = \prod_{M_l \in \mathcal{M}} \left [ \sum_{t=1}^N \left ( p_t(t) \cos^2((M_l+1)\theta_a) \right ) \right ] 
    \label{eq:QSearch_FNeg}
\end{equation}

For the current experiments let $p_t(i)=1/N$, i.e., the number of primitives potentially intersected by any ray is uniformly distributed between $1$ and $N$ ($t=0$ would not be a false negative). Equation \ref{eq:QSearch_FNeg} becomes
\begin{equation}
    \overline{p}_{QS} = \prod_{M_l \in \mathcal{M}} \left [ \frac{1}{N} \sum_{t=1}^N \cos^2((M_l+1)\theta_a) \right ] 
    \label{eq:QSearch_FNeg_Uniform}
\end{equation}

Table \ref{tab:pQS} presents these estimates for different numbers of geometric primitives (from $2^3$ to $2^9$) and $c=1.8$, which is the default value used by Qiskit.

\begin{table}[ht]
  \caption{$\overline{p}_{QS}$ - QSearch() false negative probabilities for $c=1.8$}
  \label{tab:pQS}
  \begin{minipage}{\columnwidth}
  \begin{center}
  \begin{tabular}{c|ccl|c}
\multirow{2}{*}{$N$} & $\max(M_l)  $ & \multirow{2}{*}{$L$} & $\mathcal{M}=\{ \text{round} (c^l) :$ & \multirow{2}{*}{$\overline{p}_{QS}$} \\
    & $< \left \lceil \sqrt{N} \right \rceil$ & & $l=0 \ldots L-1$ \} & \\
    \midrule
    8 & 3 & 2 & \{1, 2\} & 0.205 \\
    16 & 4 & 3 & \{1, 2, 3\} & 0.113 \\
    32 & 6 & 4 & \{1, 2, 3, 6\} & 0.055 \\
    64 & 8 & 4 & \{1, 2, 3, 6\} & 0.057 \\
    128 & 11 & 5 & \{1, 2, 3, 6, 10\} & 0.029 \\
    256 & 16 & 5 & \{1, 2, 3, 6, 10\} & 0.029 \\
    512 & 23 & 6 & \{1, 2, 3, 6, 10, 19\} & 0.015 \\
     \bottomrule
\end{tabular}
\end{center}
\smallskip
\end{minipage}
\end{table}

Table \ref{tab:ScaleNumPrims-NN-SN} presents performance and error metrics for the proposed termination criterion combined with gathering data from neighboring pixels. For the sake of clarity the number of intersections per ray is also shown for the quantum non-optimized case (taken from table \ref{tab:ScaleNumPrims}) and for the classical approach. 
\begin{table}[ht]
  \caption{Scalability with the number of primitives.}
  \label{tab:ScaleNumPrims-NN-SN}
  \begin{center}
  \begin{tabular}{rr|r|rr|rrr}
    \toprule
  &  & \multicolumn{1}{|c|}{\small Classical} & \multicolumn{5}{c}{Quantum} \\
  &  &  & \multicolumn{2}{c|}{No-Opt} & \multicolumn{3}{c}{Opt} \\
{\small Scene} & N &  \multicolumn{1}{|c|}{$\frac{\text{Int}}{\text{Ray}}$} & $\frac{\text{Int}}{\text{Ray}}$ & {\small \#Dpix} & $\frac{\text{Int}}{\text{Ray}}$ & {\small \#Dpix} & {\small \#Cpix} \\
    \midrule
     \multirow{7}{*}{\small Qornell}& 8 & 8 & 12.0 & 18 & 9.8 & 11 & 912  \\
     & 16 & 16 & 18.0 & 94 & 14.4 & 89 & 1077  \\
     & 32 & 32 & 27.4 & 93 & 21.3 & 92 & 1316  \\
     & 64 & 64 & 33.6 & 107 & 22.1 & 103 & 2438  \\
     & 128 & 128 & 50.4 & 108 & 32.6 & 108 & 2197  \\
     & 256 & 256 & 51.3 & 295 & 33.7 &  144 & 4448  \\
     & 512 & 512 & 79.7 & 157 & 51.8 & 124 & 3445 \\ \midrule
     \multirow{2}{*}{\small Depth} & 32 & 32 & 30.7 & 201 & 22.0 & 2 & 5456 \\
     & 64 & 64 & 39.4 & 95 & 22.5 & 21 & 5781 \\
\bottomrule
\end{tabular}
\end{center}
\end{table}
The following conclusions can be drawn:
\begin{itemize}
    \item the optimized self terminating approach is still $\mathcal{O}(\sqrt{N})$, but with significantly better constants than the non-optimized version. This improvement comes from the fact that rays are terminated on early iterations, thus reducing the overall workload. The number of rays active at each iteration decreases very rapidly with the iteration count, in opposition to the previous approach where all rays were traced on all iterations;
    \item the improvement on the number of intersections per ray over the non-optimized quantum version increases significantly with $N$; this hints towards much more significant gains as technology improvements allow for more complex scenes (larger $N$);
    \item there is also a gain in image quality (\#Dpix), which is particularly evident for the Depth Complexity scene (the Qornell Box has a lower bound on \#Dpix imposed by Z-fighting in the back wall edges, which is not addressed by any of these approaches). This improvement in image quality is due to the gathering of neighbouring data among pixels.
\end{itemize}

\begin{figure}[ht]
\begin{center}
    \includegraphics[width=0.45\columnwidth]{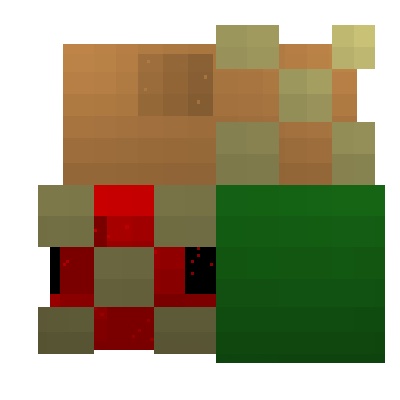}
    \caption{Depth Complexity scene with the termination criterion (64 primitives).}
\label{fig:Depth-SN-NN}
\end{center}
\end{figure}

Figures \ref{fig:Q_Qornell064_128x128_NN_SN-teaser} and \ref{fig:Depth-SN-NN} present, respectively, the rendered images for the Qornell Box and Depth Complexity scenes. 

\input{RandomClassical}

%% file: RandomClassical.tex
\subsection{Randomized Classical Algorithm}

Results from quantum algorithms are obtained through a non-unitary measurement operation, which collapses the quantum state onto a stochastically selected classical state, according to the probabilistic distribution encoded in the basis states' amplitudes. Given the stochastic behaviour of quantum algorithms it is a good practice to compare their performance against randomized classical algorithms, rather than comparing only with deterministic classical approaches. Randomized algorithms often exhibit better complexities that their deterministic equivalents.

The randomized classical renderer used within this section follows algorithm \ref{alg:TracePass} for the base version (algorithm \ref{alg:DirectPass} for shadow rays) and then algorithm \ref{alg:TracePassNN_func} for the optimized version based on gathering data from local neighbours. However, the quantum ray tracing algorithms \QTrace (algorithm \ref{alg:QTrace}) and \QOccluded are not used. Instead, a randomized classical intersection routine is used, which randomly selects $N_{sqrt} = \left \lfloor{\sqrt{N}} \right \rfloor$ geometric primitives among the total set of $N$ such primitives. The ray is tested only against this subset of primitives, thus bringing the complexity of a tracing operation down to $\mathcal{O}(\sqrt{N})$.

Applying the termination criterion (section \ref{subsec:Termination}) requires estimating the probability of a false negative for the randomized classical case, $\overline{p}_{RC}$. Let $t$ be the number of primitives intersected by a ray. A false negative is reported if none of these $t$ primitives is randomly selected into the $N_{sqrt}$ subset of primitives actually tested against the ray:  

\begin{equation}
    \overline{p}_{RC}(t) = \frac{N-t}{N} \times \frac{N-t-1}{N-1} \times \ldots \times \frac{N-t-(N_{sqrt}-1)}{N-(N_{sqrt}-1)}
    = \prod_{n=0}^{N_{sqrt}-1} \frac{N-t-n}{N-n} 
    \approx \left ( \frac{N-t}{N} \right )^{N_{sqrt}} \text{if } N \gg N_{sqrt}
\end{equation}

\begin{figure}[htp]
\begin{centering}
	\subfigure[Reference image.]{
	    \label{fig:C-Qornell064-WH128c}
        \includegraphics[width=0.37\columnwidth]{Figures/C-Qornell064-WH128_crop.jpg}
    } \hspace{0.3cm}
	\subfigure[Base version {\scriptsize (Q: fig. \ref{fig:Q_Qornell064_128x128_MP4})}.]{
	    \label{fig:RC-Qornell064-WH128_mt_6_mo_2}
        \includegraphics[width=0.37\columnwidth]{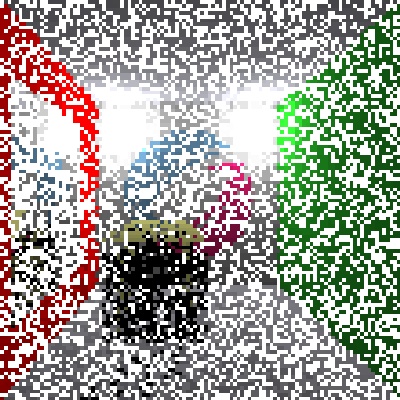}
    } \\
	\subfigure[Neighboring data {\scriptsize (Q: fig. \ref{fig:Q_Qornell064_128x128_NN_MP1})}.]{
	    \label{fig:RC-Qornell064-WH128_mt_2_mo_2_NNGather}
        \includegraphics[width=0.37\columnwidth]{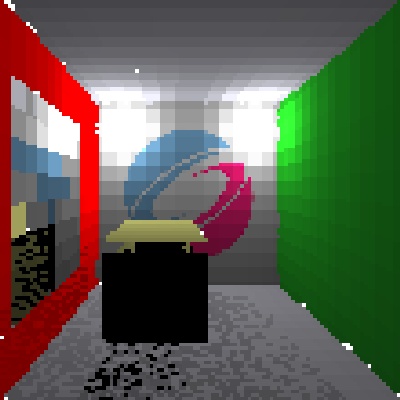}
    } \hspace{0.3cm}
	\subfigure[Termination criterion {\scriptsize (Q: fig. \ref{fig:Q_Qornell064_128x128_NN_SN-teaser})}.]{
	    \label{fig:RC-Qornell064-WH128_mo_3_NNGather_succNos}
        \includegraphics[width=0.37\columnwidth]{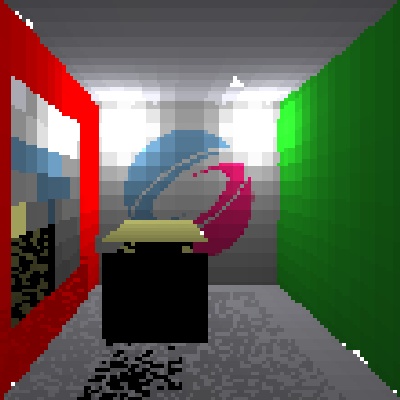}
    } 
  \caption{Randomized classical images - Qornell Box with 64 primitives. The Q in the figure captions refers to the equivalent image generated using the quantum approach.}
\label{fig:RandomClassical}
\end{centering}
\end{figure}

$\overline{p}_{RC}$ is obtained by averaging over all values of $t$ which allow for false negatives, i.e., $t=1 \ldots N-N_{sqrt}$:
\begin{equation*}
    \overline{p}_{RC} = \frac{1}{N-N_{sqrt}} \sum_{t=1}^{N-N_{sqrt}} \left ( \frac{N-t}{N} \right )^{N_{sqrt}}
\end{equation*}
For the 64 primitives scene being reported $\overline{p}_{RC} \approx 0.12$. This is significantly larger than $\overline{p}_{QS} \approx 0.057$ (see table \ref{tab:pQS}), as expected since no process similar to amplitude amplification takes place. $\overline{p}_{RC} > \overline{p}_{QS}$ implies that more \TracePass iterations are required for the randomized classical case in order to achieve identical image quality, since false negatives are more likely to occur. Or alternatively, it means that for the same computational effort, measured by the number of intersection evaluations per ray, the quantum renderer will generate higher quality images (measured by NMRSE and \#Dpix). Table \ref{tab:RandomClassical} clearly demonstrates that experimental results conform with the predicted results. For each of the three versions of the renderer (base version, exploitation of neighboring data and autonomous termination criterion) and for approximately the same ratio of intersections per ray, the quantum renderer consistently outperforms the randomized classical one. Figure \ref{fig:RandomClassical} presents the images obtained for each of these versions of the renderer -- each image caption refers to the corresponding quantum rendered image used for comparison.

\begin{savenotes}
\begin{table}[htp]
\begin{center}
  \caption{Randomized classical results: Qornell Box with 64 primitives.}
  \label{tab:RandomClassical}
  \begin{minipage}{\columnwidth}
  \begin{center}
  \begin{tabular}{l|c|ccc}
    \toprule
  \multicolumn{1}{c|}{Comment} & Alg. \footnote{Q - quantum; RC - randomized classical} & $\frac{Int}{Ray}$ & NRMSE & \#Dpix \\ \midrule
    \multirow{2}{*}{Base version}& Q & 33.6 & 4\% & 107  \\
     & RC & 33.5 & 110\% & 6296 \\ \midrule
    \multirow{2}{*}{Neighboring data} & Q &  17.4 & 0.3\% & 102  \\
     & RC & 16.9 & 7.6\% & 1325  \\ \midrule
    \multirow{2}{*}{Stop criterion}& Q & 22.1 & 0.4\% & 103  \\
    & RC & 23.4 & 7.5\% & 970  \\ \bottomrule
\end{tabular}
\end{center}
\end{minipage}
\end{center}
\end{table}
\end{savenotes}

%% file: BeyondWRT.tex
\section{Beyond Whitted ray tracing}

In this section we show further results when applying our optimized quantum algorithm to both primary and shadow rays. We illustrate this in two common contexts in rendering: direct lighting from area light sources and global illumination. In both these cases we use classical Monte Carlo sampling to generate samples, but use our hybrid quantum-classical approach to compute visibility. 

\begin{figure}[t]
\begin{centering}
	\subfigure[Classical reference: 512 light samples.]{
	    \label{fig:C-G811-WH128-NMC512}
        \includegraphics[width=0.37\columnwidth]{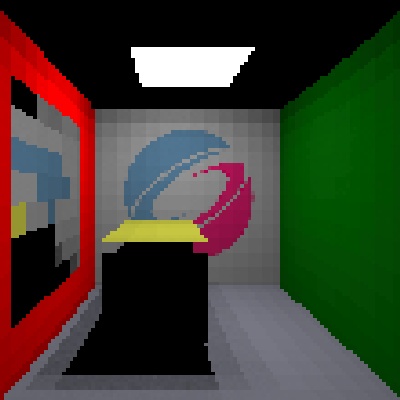}
    } \hspace{0.3cm}
	\subfigure[Quantum rendering: 32 light samples.]{
	    \label{fig:G811-WH128-NMC32}
        \includegraphics[width=0.37\columnwidth]{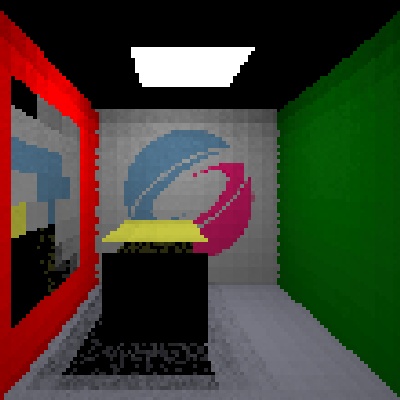}
    } \\
    \caption{Area light source lighting: Qornell Box with 64 primitives.}
\label{fig:AreaLightSource}
\end{centering}
\end{figure}

In Figure \ref{fig:C-G811-WH128-NMC512} we show results for the Qornell Box lit from an area light source in the ceiling, using 32 light samples (shadow rays) per pixel. A classically rendered image using 512 light samples is also displayed as a reference. The umbra and penumbra regions are clearly distinguishable, even though they are noisy. This noise has two causes. Firstly, a smaller number of light samples per pixel is used since computing 512 shadow rays per pixel would require unacceptable quantum simulation times. Secondly, the reuse of neighboring data optimization is not performed for shadow rays, therefore false negatives occur more frequently than for primary and specular rays. Two complementary solutions are possible: i) extend the gathering of neighboring data to shadow rays, which is planned as future work, and ii) allow \DirectPass to perform more than 2 iterations, which is straightforward but would increase simulation time. Despite the noise, these results clearly demonstrate that our hybrid approach handles area light sources, while still maintaining a quadratic complexity advantage.

\begin{figure}[t]
\begin{centering}
	\subfigure[Classical reference: 512 VPLs per pixel.]{
	    \label{fig:C_Temple_128x128_NMC_512}
        \includegraphics[width=0.37\columnwidth]{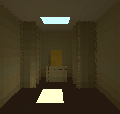}
    } \hspace{0.3cm}
	\subfigure[Quantum rendering: 4 VPLs per pixel.]{
	    \label{fig:Q_Temple_128x128_NMC_4}
        \includegraphics[width=0.37\columnwidth]{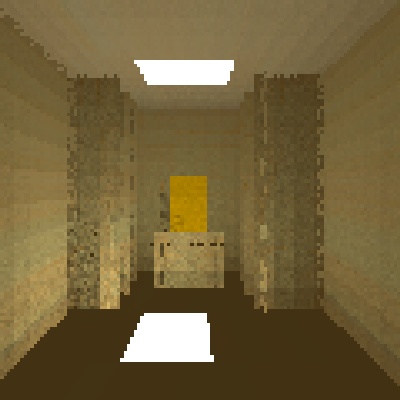}
    } \\
    \caption{Indirect lighting using single bounce VPLs per pixel}
\label{fig:IndirectLighting}
\end{centering}
\end{figure}

In Figure \ref{fig:IndirectLighting} we show results for indirect lighting in a temple scene consisting of 32 primitives. This scene is lit from directional light coming through a small hole in the ceiling. From this, single bounce indirect illumination is computed from a unique set of Virtual Point Lights (VPLs) per pixel. We use one primary ray per pixel, which leads to some noise due to Z fighting due to the integer quantization (see section \ref{sec:experimental_setup}), however the contribution of indirect lighting from the VPLs using our hybrid-classical approach for visibility in Figure \ref{fig:Q_Temple_128x128_NMC_4} clearly is approaching that of a classical reference with 512 VPLs per pixel in Figure \ref{fig:C_Temple_128x128_NMC_512}.



The number of intersections per ray using our optimizations for the area light source scene is 21.8 compared to 64 classically, and indirect lighting requires 22.4 compared to 32 classically. Performing operations classically has $\mathcal{O}(N)$ complexity as expected, with the number of intersections per ray equal to the number of primitives, whereas the hybrid quantum-classical approach leads to a quadratic improvement in the number of intersections required.


%% file: Discussion.tex
\section{Discussion}

The presented results have demonstrated a quadratic advantage of hybrid classical-quantum ray tracing over a classical approach. By exploiting amplitude amplification techniques, the core of quantum searching algorithms, the number of intersection evaluations per ray exhibits a $\mathcal{O}({\sqrt{N})}$ complexity, rather than the $\mathcal{O}(N)$ associated to classical renderers (without spatial ordering of the scene's elements). Optimizations based on exploitation of local coherence and on a principled per ray termination criterion further improve the hybrid ray tracer's performance by bringing down the constants hidden in the $\mathcal{O}({\sqrt{N})}$ expression.

The complexities reported are query complexities, meaning that the oracle is treated as a black box which efficiently implements the function of interest and whose complexity is considered to be $\mathcal{O}(1)$. It is known that if an efficient implementation of a function exists for a classical machine, then an efficient quantum implementation of the same function is also realizable with minimum overhead. If we accept that the classical function intersecting one ray with one geometric primitive is $\mathcal{O}(1)$, then the quantum oracle is also $\mathcal{O}(1)$. Under these conditions, the query complexity of the quantum program is quadratically better than that of the classical program, given that both programs have access to the same oracle.

There are however limitations of a practical nature, which might seem to compromise the quantum quadratic advantage goal. A prominent limitation is the loading of the scene's geometry data into the quantum circuit. This is currently performed by the oracle, namely by the $[m|M]_{[X|Y|Z]}$ circuits described in section \ref{subsec:IntOperator}. Being located within the oracle, our query complexity analysis treats it as a black box and assumes complexity $\mathcal{O}(1)$. However, the number of gates required to load the data might grow polynomially with the number of primitives. We use Boolean simplification techniques to reduce the number of required gates, but without any guarantee on an upper bound. In order to overcome the overhead associated with the loading of data qRAM has been proposed \cite{giovannetti2008quantum}, which would allow access to $N$ data items in $\mathcal{O}(\log(N))$ steps; qRAM is still not available, but as technology progresses it holds the potential to overcome the data loading overhead.

An additional limitation is the classical evaluation of the intersection points between the ray and the planes embedding each geometric primitive. This stage of the intersection procedure becomes $\mathcal{O}(N)$, whereas it would be $\mathcal{O}(1)$ if performed by the quantum hardware. As explained in section \ref{subsec:IntOperator}, there are no fundamental reasons why this cannot be migrated to the quantum circuit. There are practical reasons: the circuit would become so wide (number of qubits) and deep (number of gates along a critical path) that it would not map neither to a real device nor to a simulator. Identical reasons constrained the representation of all data, including the description of the 3D world, to integer values. And these are small integers, which can be represented using a small number of qubits. For instance, the 3D world is a 16x16x16 quantized volume requiring at most 4 qubits to represent a coordinate along an axis. These constraints place a hard limit on the complexity of the scenes that can be processed and produce quantization artifacts perceivable in the rendered images. However, as technology advances and quantum volume \cite{cross2019validating} increases these constraints will be lifted once floating point representations become practical.

\begin{figure}[ht]
\begin{center}
  \includegraphics[width=0.75\columnwidth]{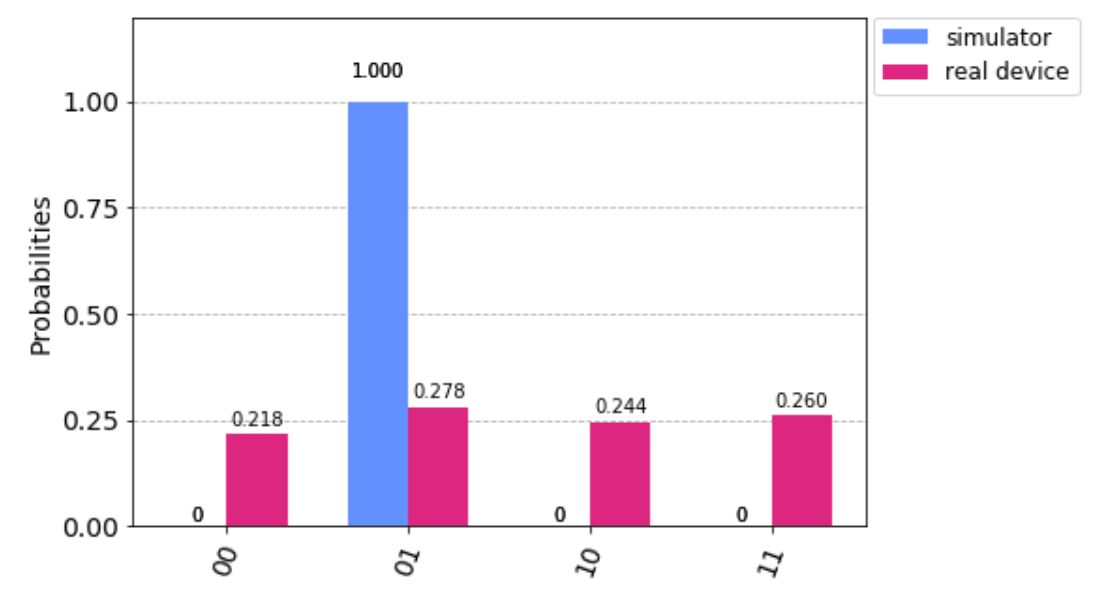}
  \caption{Simulator vs. Real Device Measurement Probabilities: 1024 shots}
\label{fig:RealDeviceHistogram}
\end{center}
\end{figure}

The above limitations allowed the proposed quantum circuits to be executed on a simulator, where results are not affected by noise associated with current NISQ systems \cite{Preskill2018}. In order to verify how do these circuits behave on a real quantum device, we designed a minimal scene with 4 geometric primitives, selected a primary ray which intersects only one of these primitives and executed the respective quantum circuit both on the simulator and on a real device. We used ibmq\_paris, a superconducting device with 27 qubits and quantum volume equal to 32. With respect to connectivity (required for multi-qubit gates), 13 out of the 27 qubits have 2 immediate neighbors, 8 have 3 neighbors and the remaining have a single neighbor. Table \ref{tab:CircuitsProperties} presents some circuit properties both for execution on the simulator and after being transpiled for execution on ibmq\_paris. A depth of $\approx 11$ K gates is at least 3 orders of magnitude larger than what can be reliably executed on current devices; this huge depth and high gate count are due to the limited connectivity of the quantum device, requiring multiple quantum state swaps such that two-qubits gates operate over neighboring qubits. Figure \ref{fig:RealDeviceHistogram} presents the histogram for the measured state after 1024 executions on each platform. Clearly, results on the real device approach an uniform distribution over all possible states, precluding an efficient use of the device for such large circuits. It is expected that technological developments will allow for a steady exponential increase in quantum volume, which has been doubling every year. Together with the advent of quantum fault tolerant systems, reliable execution of this circuit might be possible in the medium-term.
\label{subsec:RealDevice}

\begin{table}[ht]
  \caption{Quantum circuits attributes: simulator and ibmq\_paris.}
  \label{tab:CircuitsProperties}
  \begin{minipage}{\columnwidth}
  \begin{center}
  \begin{tabular}{c|ccc}
    \toprule
  Device & \#qubits & \#gates & depth  \\ \midrule
    Simulator & 19 & 2165 & 1185  \\ 
    ibmq\_paris & 27 &  18277 & 11096  \\ \bottomrule
\end{tabular}
\end{center}
\end{minipage}
\end{table}

%% file: Conclusion.tex
\section{Conclusion and Future Work}

This paper has presented the first practical hybrid quantum-classical algorithms for ray tracing. We show that compared to a classical approach, quantum algorithms improve the complexity of ray primitive intersection operations. Further to this, we presented algorithms which substantially improve performance through exploiting spatial coherence to minimize the impact of quantum randomness and through a generalised approach to a stopping criteria for the searching process. We present results for scenes with varying depth complexity, and for use with multiple light transport phenomena, such as area light sources and indirect lighting. 

We plan to extend this work in multiple directions. The first is to further investigate quantum numerical integration within our framework. This would be another step towards a fully quantum rendering system. However, as noted in section \ref{sec:RelatedWork}, this would require the quantum machine to store the entire rendering context which may require a prohibitively large amount of quantum volume. Secondly, we would like to further lift some of the limitations of this work, and as the quantum volume available on real machines increases we intend to further explore our algorithms in practice.


%% file: QuantumComputing.tex
\section{Introduction to Quantum Computing}
\label{sec:IntroductionToQuantumComputing}

The basic building block of quantum computing is a qubit $|b\rangle$, a quantum analogue of the classical bit. Rather than the classical binary values, a the quantum state of a qubit is a linear superposition of two distinct orthonormal basis states $|0\rangle = \begin{bmatrix}
1\\ 
0
\end{bmatrix}$ and $|1\rangle = \begin{bmatrix}
0\\ 
1
\end{bmatrix}$. This is described as $|b\rangle = \alpha_{0}|0\rangle + \alpha_{1}|1\rangle$, where the complex coefficients $\alpha_{0}, \alpha_{1} \in \mathbb{C}$ are referred to as amplitudes. The amplitudes' vector is required to have unit norm, i.e. $|\alpha_{0}|^{2} + |\alpha_{1}|^{2} = 1$. Upon measurement of the qubit, its quantum state will collapse onto classical state $i=0$ or $i=1$ with probability $|\alpha_{0}|^2$ and $|\alpha_{1}|^2$, respectively. Any further measurement of $|b\rangle$ will always return the same classical value.

This generalises to $n$ qubits $|b\rangle^{\otimes n} = \sum^{2^n - 1}_{i=0} \alpha_{i} |i\rangle$, where $\alpha_{i} \in \mathbb{C}$ and $\sum^{2^n - 1}_{i=0}|\alpha_{i}|^{2} = 1$ (for ease of exposition we will frequently drop the $\otimes n$ notation when describing multiple qubits). This implies that with $n$ qubits the system can simultaneously represent $N=2^{n}$ states, and any subsequent computation therefore affects these states simultaneously, the exponential quantum parallelism which gives quantum computing its advantages.

\subsection{Operations}

\begin{figure}[htb]
\centering
    \begin{equation*}
    \Qcircuit @C=1.0em @R=0.0em @!R {
        \lstick{ |0\rangle :  } & \gate{\widehat{\mathcal{H}}} & \ctrl{1} & \meter \\
        \lstick{ |0\rangle :  } & \qw & \targ & \meter \\
        & & \underbrace{}_{\widehat{\mathcal{CX}}} &
    }
    \end{equation*}
  \caption{Quantum circuit which entangles two qubits. The qubits are represented as horizontal lines across the circuit, and gates are applied to the qubits from left to right. The controlled not is explicitly highlighted here, $\bullet$ represents the control qubit and $\oplus$ represents the not operation on the target qubit. The final symbol denotes measurement.}
\label{fig:entanglement}
\end{figure}
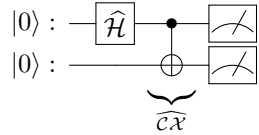

Quantum computations are frequently expressed using the quantum circuit model. Qubits are represented as horizontal lines with gates applied to them and time flowing from left to right -- see figure \ref{fig:entanglement} for an example. In this model the system is typically initialized with all qubits set to the zero state i.e. $\alpha_{0} = 1$, and then proceeds to run computations by applying operations to these qubits. These operations, or gates, are typically represented as unitary matrices $\widehat{\mathcal{U}}$ ($\widehat{\mathcal{U}}\widehat{\mathcal{U}}^{\dagger}=\widehat{\mathcal{U}}^{\dagger}\widehat{\mathcal{U}}=\widehat{\mathcal{I}}$, where $\widehat{\mathcal{I}}$ is the identity matrix) that are applied to one or more qubits. The requirement of unitary matrices implies that the unit norm of the amplitudes' vector is preserved, and that the operation is reversible. 

A common example of a one qubit gate is the Hadamard gate $\widehat{\mathcal{H}} = \frac{1}{\sqrt{2}}\begin{bmatrix}
1 & 1\\
1 & -1
\end{bmatrix}$, which is used to create superpositions. When applied to a qubit in state $|0\rangle$, an uniform superposition of the two states is created $\widehat{\mathcal{H}}|0\rangle \mapsto \frac{1}{\sqrt{2}}|0\rangle + \frac{1}{\sqrt{2}}|1\rangle$, likewise when applied to  $|1\rangle$, $\widehat{\mathcal{H}}|1\rangle \mapsto \frac{1}{\sqrt{2}}|0\rangle - \frac{1}{\sqrt{2}}|1\rangle$. This process is reversible i.e. $\widehat{\mathcal{H}}\widehat{\mathcal{H}}|0\rangle\mapsto|0\rangle$. Another common gate is the not gate $\widehat{\mathcal{X}} = \begin{bmatrix}
0 & 1\\
1 & 0
\end{bmatrix}$ which swaps the states $|0\rangle$ and $|1\rangle$ similarly to a classical not gate. A final example of a single qubit gate a phase shift gate: $\widehat{\mathcal{R(\phi)}} = \begin{bmatrix}
1 & 0\\
0 & e^{i\phi}
\end{bmatrix}$, which rotates the phase of state $|1\rangle$ by an angle $\phi$. A common specialization of this gate is the $\widehat{\mathcal{Z}}$ gate where $\phi = \pi$, and has the effect of flipping the phase of $|1\rangle$, $\widehat{\mathcal{Z}}|b\rangle \mapsto \alpha_{0}|0\rangle - \alpha_{1}|1\rangle$.

In order to perform practical computations, gates involving multiple qubits are required. The $\widehat{\mathcal{CX}}$, or controlled-not gate acts on a pair of qubits, and flips the second qubit depending of the state of the first and leaves the first unchanged, i.e. $\widehat{\mathcal{CX}}|0\rangle|0\rangle \mapsto |0\rangle|0\rangle$, $\widehat{\mathcal{CX}}|0\rangle|1\rangle \mapsto |0\rangle|1\rangle$, $\widehat{\mathcal{CX}}|1\rangle|0\rangle \mapsto |1\rangle|1\rangle$ and $\widehat{\mathcal{CX}}|1\rangle|1\rangle \mapsto |1\rangle|0\rangle$. This can be extended to multiple control qubits, known as a Toffoli gate, or generalized to any operator $\widehat{\mathcal{U}}$ conditioned on $n$ qubits, $\widehat{\mathcal{CU}} = \begin{bmatrix}
\widehat{\mathcal{I}}_{2n} & 0\\
0 & \widehat{\mathcal{U}}
\end{bmatrix}$, where $\widehat{\mathcal{I}}_{2n}$ is a $2n \times 2n$ identity matrix. This allows any conditional operation to be performed, such as conditional phase shifts \cite{Johnston2019}. When a multiple qubit operator $\widehat{\mathcal{CU}}$ is applied to a state $|b\rangle$, a relative phase of $\phi$ is induced in the control qubits $\widehat{\mathcal{CU}}(|0\rangle + |1\rangle)|b\rangle \mapsto |0\rangle|b\rangle + e^{i\phi}|1\rangle \widehat{\mathcal{U}} |b\rangle$, where $e^{i\phi}$ is an eigenvalue of $\widehat{\mathcal{U}}$, a process known as phase kickback which is exploited in many quantum algorithms.

These operators can be chained together into a program in order to perform computation. Figure \ref{fig:entanglement} shows a simple example which represents another fundamental quantum property: entanglement. The system is initialised in state $|00\rangle$. The Hadamard gate applied to the first qubit places it in a superposition,  $\frac{1}{\sqrt{2}}\left(|00\rangle + |10\rangle \right)$. Finally, a $\widehat{\mathcal{CX}}$ gate is applied to the second qubit conditioned on the first, $\frac{1}{\sqrt{2}}\left(|00\rangle + |11\rangle \right)$. 

All computations are reversible which means that information cannot be destroyed. Non reversible operations, such as the absolute value, can be transformed into reversible operations by encoding extra information into \textit{ancilla} qubits; for the absolute value example, the sign of the input can be encoded into an ancilla.

\subsection{Measurement}

At any point of the quantum circuit, the corresponding quantum state can be measured. However, this measurement returns a single classical binary state, among all those represented in the quantum superposition. State $|i\rangle$ is measured with a probability proportional to the squared absolute value of the associated coefficient, $\alpha_i$, i.e., $p(|i\rangle) = |\alpha_{1}|^{2}$. Furthermore, this measurement collapses the superposition and any subsequent measurement will return the same value. In the example shown in Figure \ref{fig:entanglement}, either $|00\rangle$ or $|11\rangle$ will be measured with probability 0.5 respectively. Note that in this example measuring the first qubit means the measurement of the second will be known; for example, if $0$ is measured for the first qubit, $0$ will be measured for the second.

%% file: QRT-algorithms-appendix.tex
\newpage

\section{Detailed hybrid rendering algorithms}
\label{sec:QRT-algs}

This presents the algorithmic implementations of the concepts described in section \ref{subsec:renderAlg}.

\subsection{DirectPass() - direct illumination}

The implementation of computing direct lighting from point light sources is presented in algorithm \ref{alg:DirectPass}.

\begin{algorithm}[H]
\SetAlgoLined
\KwData{S -- scene; rt\_maps -- rendering data}
\KwData{lights}
pixels $\leftarrow $ get\_pixels (rt\_maps) \;
pix\_light\_occ[ALL] $\leftarrow $ False \;
\tcp{visibility assessment pass}
it $\leftarrow $ 0 \;
\While{it < 2 {\bf AND} pix\_light\_occ.any()==False}{
  \ForEach{pix, lt $\in$ \{pixels $\otimes$ lights\}}{
    \If{pix\_light\_occ[pix,lt] == False}{
      ray $\leftarrow $ get\_shadow\_ray (pix, lt) \;
      occ = QOccluded (ray, S) \; 
      \lIf{occ}{
        pix\_light\_occ[pix,lt] == True 
      }
    }
  }
  it = it + 1 \;
}
\Return pix\_light\_occ
\caption{DirectPass(S, rt\_maps, lights)}
\label{alg:DirectPass}
\end{algorithm}

\subsection{RenderScene() - hybrid renderer}

Algorithm \ref{alg:RenderScene} presents the overall algorithmic approach towards rendering the scenes using a series of passes.

\begin{algorithm}[H]
\SetAlgoLined
rt\_maps $\leftarrow $ [] \;
\tcp{primary pass}
pixels $\leftarrow $ get\_pixels (cam) \;
Prays $\leftarrow $ get\_rays (pixels) \;
rt\_maps $\leftarrow $ TracePass (pixels, Prays, scene, \#IT, rt\_maps) \;
\tcp{specular pass}
Spixels $\leftarrow $ get\_spec\_pixels (rt\_maps) \;
Srays $\leftarrow $ get\_spec\_rays (Spixels) \;
rt\_maps $\leftarrow $ TracePass (Spixels, Srays, scene, \#IT, rt\_maps) \;
\tcp{direct light pass}
pix\_light\_occluded $\leftarrow $ DirectPass (scene, rt\_maps, lights) \; \label{algline:DirectPass}
\tcp{shading pass}
\ForEach{pix, lt $\in$ \{pixels $\otimes$ lights\}}{
  \If{pix\_light\_occluded[pix,lt] == False}{
    image[pix] += shade (pix, lt, scene, rt\_maps) \;
  }
}
\Return image
\caption{RenderScene()}
\label{alg:RenderScene}
\end{algorithm}